\documentclass[times]{nmeauth}

\usepackage{amssymb}
\usepackage{amsmath}

\def \Rdof {{\mathrm{dof}}}
\def \RLIN {{\mathrm{LIN}\,}}

\def \Rmod {{\mathrm{mod}}}
\def \Rnd {{\mathrm{nd}}}
\def \Rend {{\mathrm{end}}}
\def \ds {\,\Rd s}
\def \dt {\,\Rd t}

\def \dsb {\,\Rd \bar{s}}
\def \sb {\bar{s}}
\def \su {s_\Ru}

\def \dx    {\,\Rd x}
\def \dy    {\,\Rd y}

\def \dO    {\,\Rd\OM}

\def \ds    {\,\Rd s}

\def \pt  {{\partial}}

\newcommand{\fracpt}[2]{\frac{\pt #1}{\pt #2}}
\newcommand{\fracdif}[2]{\frac{\mathrm{d} #1}{\mathrm{d} #2}}

\def \Rd  {{\mathrm{d}}}

\def \Ri  {{\mathrm{i}}}

\def \Rk  {{\mathrm{k}}}

\def \Rm  {{\mathrm{m}}}
\def \Rn  {{\mathrm{n}}}
\def \Ro  {{\mathrm{o}}}
\def \Rp  {{\mathrm{p}}}

\def \Rt  {{\mathrm{t}}}
\def \Ru  {{\mathrm{u}}}

\def \RA  {{\mathrm{A}}}
\def \RB  {{\mathrm{B}}}

\def \RE  {{\mathrm{E}}}

\def \RT  {{\mathrm{T}}}

\def \Rint    {{\mathrm{int}}}

\def \Rele    {{\mathrm{ele}}}

\def \Rmin    {{\mathrm{min}}}

\def \Rnd     {{\mathrm{nd}}}

\def \Rend    {{\mathrm{end}}}

\def \BB      {{\mathbf B}}
\def \BC      {{\mathbf C}}
\def \BD      {{\mathbf D}}
\def \BE      {{\mathbf E}}
\def \BF      {{\mathbf F}}

\def \BJ      {{\mathbf J}}
\def \BK      {{\mathbf K}}

\def \BN      {{\mathbf N}}

\def \BR      {{\mathbf R}}
\def \BS      {{\mathbf S}}

\def \BX      {{\mathbf X}}

\def \Bd      {{\mathbf d}}

\def \Bs      {{\mathbf s}}

\def \Bu      {{\mathbf u}}

\def \Bx      {{\mathbf x}}
\def \By      {{\mathbf y}}

\def \de      {{\delta}}

\def \DE      {{\Delta}}

\def \PI      {{\Pi}}

\def \OM      {{\Omega}}

\def \Bxi   {{\boldsymbol \xi}}

\def\volumeyear{2020}
\begin{document}

\runningheads{R.~Sachse and M.~Bischoff}{A variational formulation for motion design of structures}

\title{A variational formulation for motion design of adaptive compliant structures}

\author{Renate~Sachse\affil{1}$^,$\corrauth and Manfred~Bischoff\affil{1}}

\address{\centering
\affilnum{1}\ Institute for Structural Mechanics, University of Stuttgart, \\ Pfaffenwaldring 7, D-70550 Stuttgart, Germany}

\corraddr{Renate~Sachse, Institute for Structural Mechanics, University of Stuttgart, Pfaffenwaldring 7, D-70550 Stuttgart, Germany, E-mail: sachse@ibb.uni-stuttgart.de}

\begin{abstract}
Adaptive structures are characterized by their ability to adjust their geometrical and other properties to changing loads or requirements during service.
This contribution deals with a method for the design of quasi-static motions of structures between two prescribed geometrical configurations that are optimal with regard to a specified quality function while taking large deformations into account. It is based on a variational formulation and the solution by two finite element discretizations, the spatial discretization (the standard finite element mesh) and an additional discretization of the deformation path or trajectory. For the investigations, an exemplary objective function, the minimization of the internal energy, integrated along the deformation path, is used. The method for motion design presented herein uses the Newton-Raphson method as a second order optimization algorithm and allows for analytical sensitivity analysis. The proposed method is verified and its properties are investigated by benchmark examples including rigid body motions, instability phenomena and determination of inextensible deformations of shells.
\end{abstract}


\maketitle

%%%%%%%%%%%%%%%%%%%%%%%%%%%%%%%%%%%%%%%%%%%%%%%%%%%%%%%%%%%%%
\section{Introduction}
\label{sec:intro}

Energy efficiency and sustainability play an increasing role in engineering and architecture. Reducing the amount of material used for construction does not only save resources but also reduces embedded energy. One possibility to realize extreme lightweight design is to make use of adaptive structures that optimally adjust their geometry to current and changing conditions by active motion. Here, two fundamentally different types of adaption via geometry change can be distinguished.

The first type of adaptive structure serves the purpose of adapting to changing loads or requirements and is mostly referred to as a smart structure.
%Conservative structures are designed to guarantee the ultimate limit state or serviceability for the ``worst case'' load scenario, leading to extensive use of material to provide the needed strength and stiffness.
A system of sensors and actuators allows the structure to adjust, respond and actively counteract to varying loads. Thus, stresses, deformations or vibrations may be reduced in order to ensure usability and serviceability~\cite{housner_structural_1997,sobek_adaptive_2001,spencer_b._f._state_2003,korkmaz_review_2011}. This approach potentially allows for eminent material savings. Examples for this type of adaptive structure are the Smart Shell~\cite{neuhauser_stuttgart_2013} or an ``infinitely stiff'' cantilever beam~\cite{senatore_shape_2017} as well as a highrise building, which serves as a demonstrator in a current research project dealing with such adaptive structures~\cite{weidner_implementation_2018}. The shape transition is realized by different types of actuators, from piezoelectric or electroactive polymer actuators~\cite{irschik_review_2002,irschik_use_2010,bar-cohen_electroactive_2019}, up to shape-memory alloys~\cite{mohd_jani_review_2014}, which are usually controlled by optimal control algorithms~\cite{preumont_vibration_2011}. The aforementioned types of adaptive structures are characterized by the fact that the individual states differ only by minor changes in geometry such that geometrically linear analyses are sufficient.

The second type of adaptive structure is not intended to adapt to varying loads but to changing requirements during service. This is, for example, the case for deployable and retractable structures. Prominent examples are the opening and closing of roofs, especially of stadiums (e.g. the Commerzbank-Arena in Frankfurt, Germany~\cite{goppert_spoked_2007}) or adaptive folding bridges (e.g. the Kiel Hörn Footbridge in Kiel, Germany~\cite{knippers_folding_2000}). But also adaptive façade elements, beyond conventional sun-blinds, which can be opened and closed depending on the position of the sun and the control of interior daylight, belong to this kind of adaptive structures and may significantly contribute to the energy efficiency of a building~\cite{grosso_adaptive_2010}. Realisations are e.g. the One Ocean Expo 2012 Pavilion in Korea~\cite{knippers_bio-inspirierte_2013} or the biomimetic façade elements Flectofin~\cite{lienhard_flectofin:_2011} and Flectofold~\cite{korner_flectofoldbiomimetic_2018}. Another current research field, where this type of adaptive structure plays a prominent role, is the shape change of morphing wings of airplanes~\cite{maute_integrated_2006,liebe_shape-adaptive_2006,vasista_realization_2012,weisshaar_morphing_2013,ajaj_morphing_2016}.

In this case, the geometries of the individual configurations differ significantly from each other. The standard approach to achieve variability in geometry is the targeted introduction of joints and hinges between stiff elements and associated defined kinematics by unfolding, sliding and similar mechanisms. But particularly these joints and hinges frequently represent weak spots of the structure and may be prone to failure. Another strategy for geometrical variability are discrete systems with integrated actuators. Such systems, which can also account for large deformations, can take the form of trusses~\cite{balaguer_development_2008,sofla_shape_2009,senatore_synthesis_2019}, tensegrity structures~\cite{graells_rovira_control_2009,wijdeven_shape_2005,masic_path_2005,veuve_adaptive_2017,sychterz_deployment_2018} or lattice structures~\cite{friedman_overview_2013}, where individual actuator elements can vary their length and therefore change the shape of the entire structure. This strategy is in contrast to an overall flexibility that is distributed in the entire structure and is therefore able to perform a smoothly distributed motion. This is, for example, the case for pure bending deformations, so-called inextensional deformations, in flexible and shape changing shells~\cite{pagitz_shape-changing_2013,bigoni_folding_2015}. There are also approaches available to combine the discrete flexibility by joints with a distributed structural flexibility. Recent research investigates the optimization of flexibility and compliance of structures in order to enable an efficient deformation. These so-called compliant or morphing structures~\cite{sigmund_design_1997,saggere_static_1999,kota_design_2001,lu_design_2003,lu_effective_2005,campanile_modal_2008,hasse_design_2009,masching_parameter_2016,geiser_gacm_2017} are characterized by continuous stiffness changes and a varying stiffness distribution within the structure, leading to the formation of specific hinge zones. The challenge here is that, despite compliance, the structure remains strong enough to still be able to withstand loads in all configurations. The concept of multistable compliant structures also represents a possibility to deal with this problem and, at the same time, to keep the configurations stable without continuously expending effort~\cite{oh_synthesis_2009,santer_compliant_2008}. 

Not only the geometries of the individual configurations at the beginning and the end of the motion have to meet specified requirements, but also the shape transition between the configurations. These transitions and movements imply stress onto the structure and require energy. The way of controlling the actuators plays a decisive role in the efficiency of the morphing structure.

In control theory, especially in optimal control and in motion planning of robots~\cite{elbanhawi_sampling-based_2014,lavalle_planning_2006}, exactly this problem of optimal trajectories is already addressed. Most structures that are investigated and calculated in these research areas, e.g. robots, are characterized by a discrete kinematic description and no (or negligible) elastic deformation. This leads to fewer degrees of freedom compared to continuously morphing structures. However, in the field of continuum robotics and hyper-redundant manipulation, exactly such continuous robots or systems are planned and investigated. A review of this field is given in Rus and Tolley~\cite{rus_design_2015}. But also in this case, consideration of a large number of degrees of freedom in combination with motion planning and optimal control strategies is challenging. Therefore, simplifications are made to capture the kinematics, like a piecewise constant curvature (PCC) model~\cite{webster_design_2010}. This enables control of such systems and solution of the inverse kinematics problem (calculating the required curvatures for a given end position), but planning and optimal control methods for continuous robots without a PCC model still remains a challenge and an open research tasks. 

There are already examples and methods where the mechanics and analysis of structures are combined with optimal control and motion planning strategies, especially with regard to adaptive structures and large deformations. Ibrahimbegovic et al.~\cite{ibrahimbegovic_optimal_2004} successfully combined optimal control with non-linear structural mechanics of a beam to reach a deformed end configuration with certain properties. Furthermore, Veuve et al.~\cite{veuve_adaptive_2017}, Sychterz and Smith~\cite{sychterz_deployment_2018} and Masic and Skelton~\cite{masic_path_2005} used motion planning algorithms for the trajectory planning of tensegrity structures.

The focus of this study, however, is analysis and optimization with respect to certain requirements of the trajectory itself between prescribed configurations of any kind of flexible structure without a control algorithm and related sensoric equipment. In Werter et al.~\cite{werter_two-level_2013}, the required actuation energy was already considered for the design of a morphing airplane wing and Maute and Reich~\cite{maute_integrated_2006} combined geometry optimization of a compliant wing structure with additional optimization of the adaption mechanism. Also, the pure transition between configurations can be formulated as an optimization problem. However, when non-linear kinematics is considered any evaluation of the objective function consists of a complete non-linear analysis, including all pertinent issues like convergence problems, instabilities and high computational cost. The use of higher order optimization methods requires fewer evaluations of the objective function but the required sensitivity analyses again cause computational expense.

This contribution deals with shape transitions as a motion between two (or more) geometrical configurations, taking geometrically non-linear structural behavior into account. The idea is to design shape transitions based on a variational formulation. A quasi-static process is assumed such that no inertia effects are considered. The solution is found by a finite element discretization of the path (trajectory) along with a Newton-Raphson solution algorithm, which can be interpreted as a second order optimization algorithm. Due to the path discretization, analytical sensitivities can be calculated by making use of standard components of the spatial finite elements, e.g. the stiffness matrix.

The paper is organized as follows. First, we refer to the Brachistochrone problem, which represents one of the first problems solved by variational principles, as an illustrative example to motivate the use of a variational method for motion design. Chapter~\ref{sec:brachistochrone} presents the problem statement and its solution with different strategies. In Section~\ref{sec:motion_design}, a finite element solution algorithm is described to solve the problem of motion design of structures. The underlying weak form is based on a functional to be minimized to obtain an ``optimal'' path. The method is developed with the exemplary objective of minimizing the integral of the strain energy along the entire motion path. By a discretization of the motion path with finite elements, in addition to spatial discretization, a non-linear system of equations is obtained. Chapter~\ref{sec:num_exp} presents several numerical experiments to verify the proposed method by means of problems with known exact solutions, for instance, the motion of a kinematic system with zero strain energy throughout the entire process. Furthermore, motions dominated by instability behavior (following the motivation by~\cite{reis_perspective_2015,hu_buckling-induced_2015}) and further potential of motion design and applications for a number of flexible structures, like the calculation of inextensible deformations of shells, are investigated. Finally, some conclusions are given and open issues and potential future developments are discussed in Chapter~\ref{sec:conclusions}.

\section{The Brachistochrone problem}
\label{sec:brachistochrone}

\subsection{Historical background}

To motivate the use of a variational formulation for the solution, the classical Brachistochrone problem is considered and similarities to the problem of motion design are highlighted. The Brachistochrone problem is one of the first problems that was solved with the calculus of variations and it represents the base for its development. In 1696, in the journal ``Acta Eruditorum'', published by Gottfried Wilhelm Leibniz, Johann Bernoulli posed to the scientific community the following challenge:

\begin{quote}
\emph{Given two points A and B in a vertical plane, what is the curve traced out by a point acted on only by gravity, which starts at A and reaches B in the shortest time?} \cite{bernoulli_1696}
\end{quote}

Shortly after the publication of the problem, Bernoulli received a letter from Leibniz in which he explained that he ``is attracted by the problem like Eva by the apple'', but at the same time, he asked for an extension of the deadline, since the problem reached other countries only after a few months. Bernoulli agreed with the proposal and reformulated the problem. At the beginning of 1697, an anonymous solution to the problem appeared in the journal ``Philosophical Transactions of the Royal Society of London''~\cite{newton_1697} and finally in May 1697 Leibniz published a collection of the submitted solutions~\cite{leibniz_1697}, in which also the anonymous solution, which Bernoulli identified directly as Newton's solution (``from the claw of the lion''), was reprinted. The solution by Jacob, Johann's brother, was then further developed and a few years later Leonhard Euler named it the calculus of variations~\cite{euler_var}.

\subsection{Solution to the Brachistochrone problem}

\subsubsection{Derivation of the required time as a functional}

\begin{figure}[h]
\centering
\includegraphics[width=1.0\textwidth]{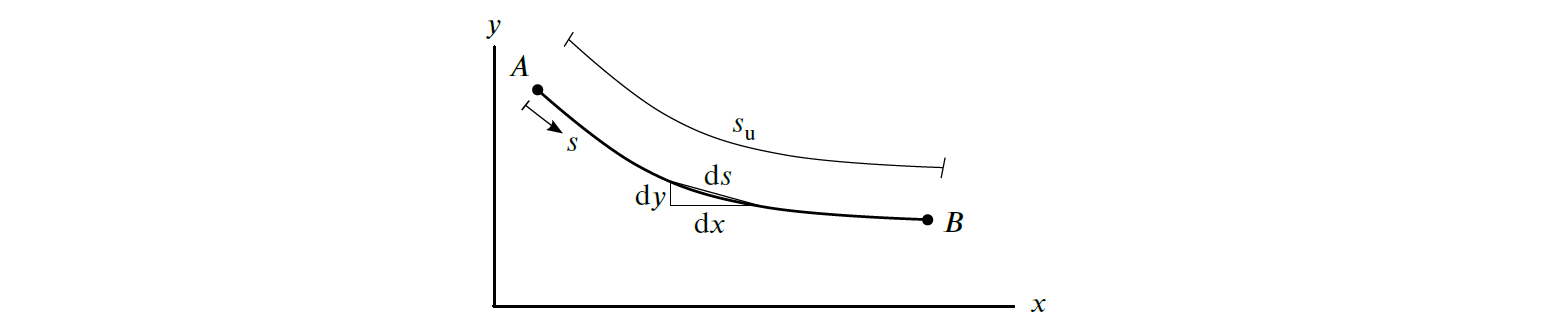}
\caption{Brachistochrone problem}
\label{fig:brachistochrone_problem}
\end{figure}

Starting point for the calculation is conservation of energy with the kinetic energy $E_{\Rk\Ri\Rn}$ and the potential energy $E_{\Rp\Ro\Rt}$
\begin{align}
E_{\Rk\Ri\Rn} + E_{\Rp\Ro\Rt} = \frac 12 mv^2 +mg y(x) = mgy_\RA,
\end{align}
where $m$ represents the mass, $v$ the velocity, $y$ is the vertical abscissa and $mgy_\RA = \mathrm{const.}$ defines the reference energy. Solving for the velocity yields
\begin{align}
v = \sqrt{2g \big(y_\RA - y(x)\big)}.
\end{align} 
By using the definition of the velocity as time derivative of the arc length $v = \fracdif{s}{t}$ , an infinitesimal time increment can be written as $\Rd t = \frac{1}{v} \ds$. The total time required for traveling from A to B is thus
\begin{align}
T = \int_0^T \dt = \int_{s_\RA}^{s_\RB} \frac{1}{\sqrt{2g \big(y_\RA - y(x)\big)}} \ds = \Rmin .
\label{eq:brach_time}
\end{align}
The infinitesimal arc length $\ds$ can be calculated from the Pythagorean theorem,
\begin{align}
\ds = \sqrt{\dx^2+\dy^2} = \sqrt{\dx^2+\dy^2} \frac{\dx}{\dx} = \sqrt{\Big(\frac{\dx}{\dx}\Big)^2+\Big(\frac{\dy}{\dx}\Big)^2} \dx = \sqrt{1+ y'(x)^2} \dx.
\label{eq:brach_arclength}
\end{align}
An illustration for this derivation is given in Figure~\ref{fig:brachistochrone_problem}.
Combining eq.~(\ref{eq:brach_time}) and eq.~(\ref{eq:brach_arclength}) yields the functional for the Brachistochrone problem
\begin{align}
T & = \int_{s_\RA}^{s_\RB} \frac{1}{\sqrt{2g \big(y_\RA - y(x)\big)}} \ds
= \int_{x_\RA}^{x_\RB} \sqrt{\frac{1+y'(x)^2}{2g\big(y_\RA - y(x)\big)}} \dx .
\label{eq:brach_func}
\end{align}

\subsubsection{Euler Lagrange equation and exact solution}
\label{sec:brachistochrone_exact}

As the integrand $F$ of the functional $T$ (eq.~\ref{eq:brach_func}) depends on the function $y$ and its derivative $y'$ only and beyond this not explicitly on the variable $x$ itself, the simplified Euler-Lagrange equation \cite{elsgolc_1970}
\begin{align}
F-y'\fracpt{F}{y'} = C
\label{eq:euler-lagrange}
\end{align}
can be used. Application of eq.~(\ref{eq:euler-lagrange}) to the functional of the Brachistochrone problem yields
\begin{align}
& \sqrt{\frac{1+y'^2}{2g(y_\RA - y)}} - y'\frac{y'}{\sqrt{2g(y_\RA-y)(1+y'^2)}}
= \frac{1}{\sqrt{2g(y_\RA-y)(1+y'^2)}} = C_1 .
\end{align}
A substitution is used for solution of the problem, where trigonometric functions and a parametric form turn out to be a clever choice:
\begin{align}
2gC_1^2(y_\RA-y) = \sin^2\left(\frac{\bar{t}}{2}\right) = \frac 12 \big(1-\cos(\bar{t})\big).
\label{eq:subst}
\end{align}
The complete derivation is omitted at this point but can be looked up in the appendix in Chapter~\ref{app:brachistochrone_exact}. Eventually, a parametric representation of the coordinates $x$ and $y$is obtained that is the function of a cycloid
\begin{align}
y(\bar{t}) & = y_\RA - \frac{1}{4gC_1^2}\big(1-\cos(\bar{t})\big) &
x(\bar{t}) & = \frac{1}{4gC_1^2} \left( \bar{t}- \sin (\bar{t}) \right) + C_2 .
\end{align}
It must be noted that $\bar{t}$ does neither represent the time nor the arc length, but is the angle by which a rolling circle has rotated, a point of which generates the curve $(x(\bar{t}),y(\bar{t}))$. The constants $C_1$ and $C_2$ as well as the parameter value $\bar{t}_\RE$ at point B are derived by the boundary conditions at the starting point A and the endpoint B: $x(\bar{t}=0) = x_\RA$, $x(\bar{t}=\bar{t}_\RE) = x_\RB$ and $y(\bar{t}=\bar{t}_\RE) = y_\RB$, which themselves represent non-linear functions that need to be solved iteratively. The condition $y(\bar{t}=0) = y_\RA$ is fulfilled by definition of the problem. An exemplary solution with fixed points A and B is given in Figure~\ref{fig:brachistochrone_solution} (left).

\subsubsection{Solution with finite elements}

In general variational problems, a closed form solution is often not available. In such cases, an approximate solution may be obtained by the finite element method. This requires the functional to be formulated in a parametric form
\begin{align}
T & = \int_{s_\RA=0}^{s_\RB} \frac{1}{\sqrt{2g \big(y_\RA - y(s)\big)}} \ds ,
\end{align}
where $s$ represents a path parameter. As the total length of the solution curve is initially unknown, the integration bound $s_\RB$ is unknown as well. Therefore, a mapping parameter $\su$ is defined, enabling integration over a variable $\sb$ and a fixed and specified domain $\sb \in [0,1]$
\begin{align}
\int_0^{s_\RB} (\ldots) \ds = \int_0^1 (\ldots) \frac{\ds}{\dsb} \dsb = \int_0^1 (\ldots) \su \dsb .
\end{align}
The mapping parameter contains information about the arc length itself
\begin{align}
\su := \frac{\ds}{\dsb} = \frac{\sqrt{\dx^2 + \dy^2}}{\dsb} = \sqrt{\left(\frac{\dx}{\dsb}\right)^2 + \left(\frac{\dy}{\dsb}\right)^2} =\sqrt{x'(\sb)^2 + y'(\sb)^2}.
\end{align}
Inserting this into the functional yields
\begin{align}
T & = \int_{0}^{1} \sqrt{\frac{x'(\sb)^2 + 
y'(\sb)^2}{2g\big(y_\RA - y(\sb)\big)}} \dsb
\end{align}
Variation with respect to the unknown functions $x$ and $y$ then follows as
\begin{align}
\de T & = \int_{0}^{1} \Bigg( \sqrt{ \frac{x'^2 + y'^2}{ 8g\big(y_\RA-y\big)^3}} \de y + \frac{x'\de x' + y'\de y'}{\sqrt{2g\big(y_\RA - y\big)\big(x'^2 + y'^2\big)}} \de x' \Bigg) \dsb = 0 .
\end{align}
This is the weak form of the Brachistochrone problem. The next steps follow the standard procedure of a finite element formulation. First, a discretization for $x$ and $y$ as well as their variations $\de x$ and $\de y$ is introduced
\begin{align}
x \approx x_h = \BN \Bx \qquad
\de x \approx \de x_h = \BN \de \Bx \qquad
y \approx y_h = \BN \By \qquad
\de y \approx \de y_h = \BN \de \By
\end{align}
where the matrix $\BN$ contains the shape functions and the vectors $\Bx$, $\By$, $\de\Bx$ and $\de \By$ contain discrete nodal values of the unknowns $x$ and $y$, respectively.
In this case, the discretization is the same for every function, following the Bubnov-Galerkin approach. By inserting the discretization into the variation
\begin{align}
\de T & = \int_{0}^1 \Bigg( \sqrt{ \frac{ (\BN'\Bx)^2 + (\BN'\By)^2}{ 8g\big(y_\RA-(\BN\By)\big)^3}} \BN \de\By + \frac{\BN'\Bx \BN' \de \Bx + \BN'\By\BN' \de \By}{\sqrt{2g\big(y_\RA - (\BN\By))\big)\big((\BN'\Bx)^2 + (\BN'\By)^2\big)}} \Bigg) \dsb = 0,
\end{align}
moving the vectors $\de \Bx$ and $\de \By$ out of the integral and applying the discrete form of the fundamental lemma of the calculus of variations, a residual is obtained. After linearization it can be solved iteratively for the nodal values $\Bx$ and $\By$, which provide an approximation for the solution functions $x$ and $y$ in a parametric form. This discretization represents a discretization of the path that the point with mass $m$ follows from A to B and is therefore referred to as path discretization in the following.

\begin{figure}[b]
\centering
\includegraphics[width=1.0\textwidth]{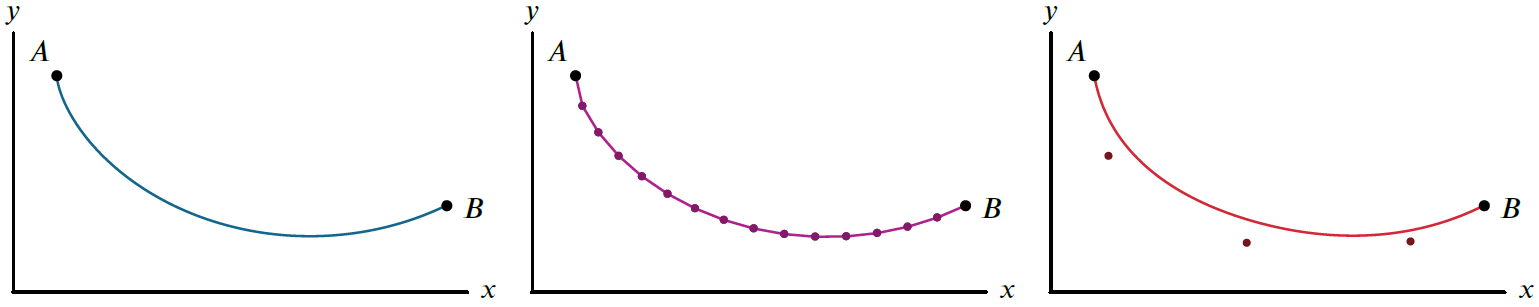}
\caption{Exact solution (left), solution with finite elements by a linear (center) and a cubic B-spline discretization (right)}
\label{fig:brachistochrone_solution}
\end{figure}

Figure~\ref{fig:brachistochrone_solution} (center) shows the result with a path discretization with linear Lagrange shape functions and 15 elements. As the parametrization of the curve, and thus the placement of the nodes along the solution curve, is not unique, an equal length of the path elements is enforced for the regularization of the solution. This extra constraint is enforced by Lagrange multipliers. It can be seen that the solution with finite elements and linear shape functions approximates well the exact curve, obtained in \ref{sec:brachistochrone_exact}, but, due to the linear functions, it still contains kinks. The number of degrees of freedom is twice the number of internal nodes ($x$- and $y$-coordinate at each node). An improvement of the approximation is possible by using an interpolation by B-spline-functions, which can also be seen in Figure~\ref{fig:brachistochrone_solution} (right), where the solution for a discretization with two elements and cubic shape functions is illustrated. The higher continuity enables a better approximation without kinks and fewer internal nodes (or control points) resulting in fewer degrees of freedom.

This problem formulation of the Brachistochrone can be taken as a simplified template for what is intended to be done in motion design. The goal is finding a path between two configurations, e.g. the points A and B or an open and a closed geometry of an adaptive element that fulfills specified demands like minimization of the total required time, energy or effort for traversing from A to B, or any other objective.

\section{Motion design of structures}
\label{sec:motion_design}

\subsection{Functional and its variation as the starting point for motion design}

\subsubsection{Integrated internal energy as objective function}

The objective of motion design is mathematically expressed as minimization of a functional. In the Brachistochrone problem this was the required time for the mass point to run from A to B. In motion design, it can be any property that the motion is expected to have. One objective could be to minimize the straining that a structure is subjected to while undergoing a certain motion. This is comparable to the dimensionless quantity \textit{cost of transport} that is used in various disciplines like biology and robotics. It represents a measure to quantify the cost or energy efficiency of different transport methods, i.~e. walking, swimming or flying of an animal or driving of a vehicle from one location to an other. Here, the notion is transferred to a \textit{cost of deformation} for flexible structures, where an energy criterion is utilized. Thus, the internal energy integrated over the entire motion path $s$ is chosen here as an exemplary functional that represents the strain energy integrated along the motion path: 
\begin{align}
J &= \int_s \PI_\Rint \ds = \int_s \int_\OM \frac 12 \BE^\RT \BS \dO \ds = \int_s \int_\OM \frac 12 \BE^\RT \BC \BE \dO \ds = \min.
\label{eq:functional}
\end{align}
Assuming small strains (but large displacements and rotations), a linear elastic St.~Venant-Kirchhoff material law is used for the relationship between Green-Lagrange strain and second Piola-Kirchhoff stress. Incidentally, it is pointed out that this functional serves as a proof of concept and can be replaced by other objectives.
\begin{figure}[b]
\centering
\includegraphics[width=1.0\textwidth]{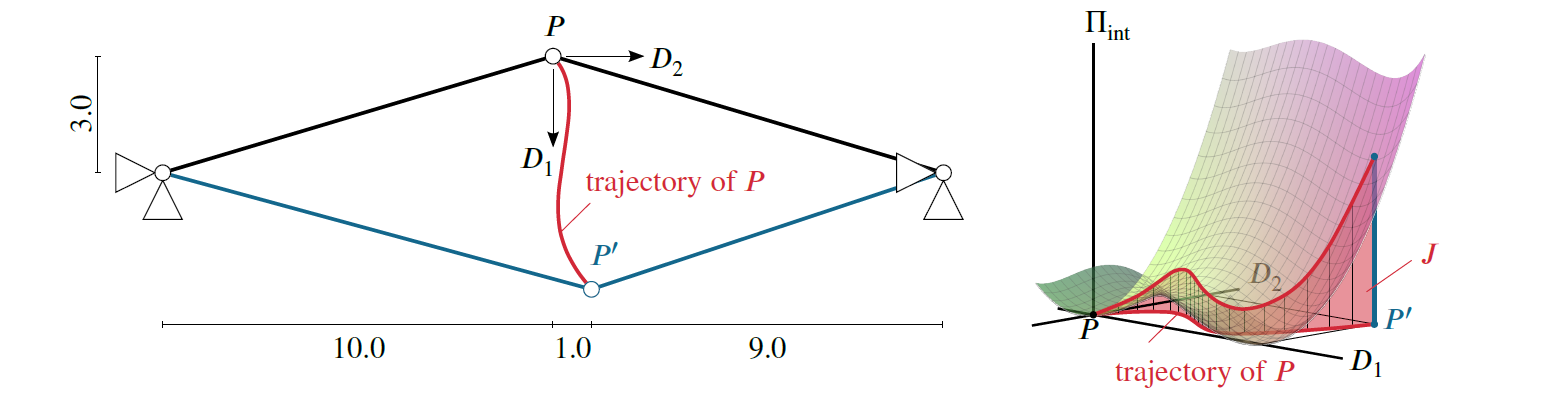}
\caption{Two bar truss with initial configuration, target configuration and visualization of the functional}
\label{fig:example_functional}
\end{figure}

For further explanation of the problem and objective, an illustrating example is introduced in Figure~\ref{fig:example_functional}. A simple truss structure (Young's modulus $E=30000$, cross section area $A=0.1$), forming a shallow arc, is supposed to deform from a starting configuration, shown in black, to a target configuration, shown in blue. This scenario is obviously inspired by the bi-stable setup of a snap-through problem. The blue target configuration, however, is not the stress-free snapped-through configuration of the black one but deviates from it by a horizontal shift of the central node.

A motion that minimizes the integrated internal energy, i.e. the functional, is to be found. The structure contains two unconstrained displacement degrees of freedom and it is assumed that forces can be applied to both of them to reach the target configuration. During deformation, the point $P$ follows the, yet unknown, trajectory (red) until it arrives at the end position $P'$. This trajectory can also be found on the plane which is spanned by the axes $D_1$ and $D_2$ in the diagram on the right in Figure~\ref{fig:example_functional}. On the vertical axes, the internal energy $\PI_\Rint$ throughout the motion is plotted as a curved line, lying on the corresponding potential function. The area of the resulting surface is the value of the functional. The goal of this specific motion design task is to find the trajectory (the motion) that minimizes this area.

\subsubsection{Specification of motion and arc length}
\label{sec:arclength}

\begin{figure}[h]
\centering
\includegraphics[width=1.0\textwidth]{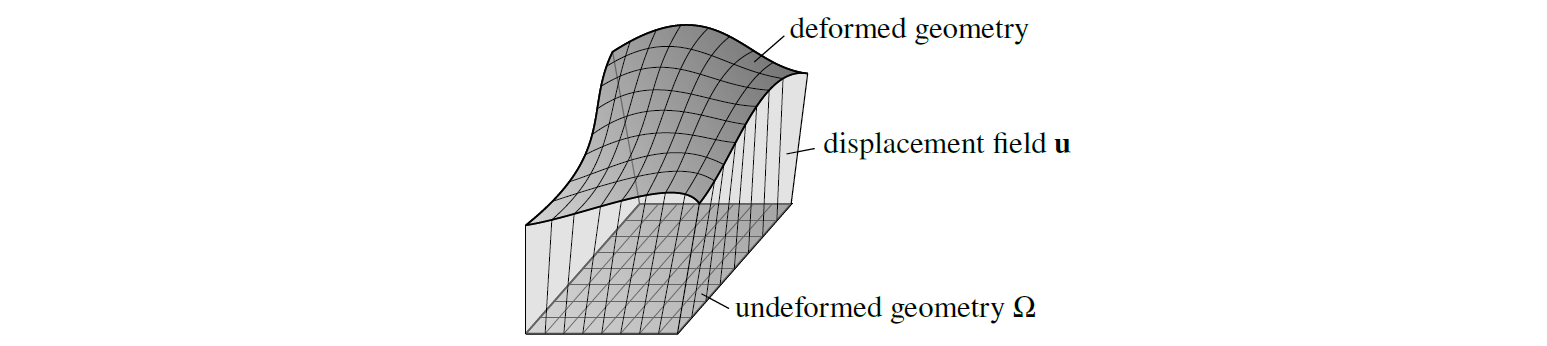}
\caption{Illustration of the displacement field}
\label{fig:arclength}
\end{figure}

The length of the path, along which the internal energy of the structure is integrated, can be associated with the arc length of the displacement field (cf. Figure~\ref{fig:arclength}) of the underlying motion, which in turn is a function of the position $\BX$ of the structure as well as the progress of the motion, the pseudo time $t$
\begin{align}
\Bu (\BX,t) = \begin{bmatrix} u_1(\BX,t) \\ u_2(\BX,t) \\ u_3(\BX,t) \end{bmatrix}.
\end{align}
In order to consider the motion in its entirety within the functional, the internal energy is integrated along the deformation path $s$. This deformation path $s$ represents a scalar measure that indicates by how much the structure has already moved and deformed. It is defined here as the arc length of the displacement field $\Bu(\BX,t)$. To obtain one scalar quantity from the displacement field, depending on the position vector $\BX$ (Figure~\ref{fig:arclength}), the mean value of the displacement arc length inside the spatial domain $\OM$ is used. Based on the same derivation as in the Brachistochrone problem (see Figure~\ref{fig:brachistochrone_problem} and eq.~(\ref{eq:brach_arclength}), an infinitesimal arc length can then be specified for a three-dimensional problem as
\begin{align}
\ds = \frac{1}{V} \int_\OM \sqrt{\Rd u_1^2 + \Rd u_2^2 + \Rd u_3^2} \dO
\end{align}
including the volume $V$ of the domain. In the illustrating example of Figure~\ref{fig:example_functional}, the arc length turns out to be the length of the trajectory of point $P$ multiplied with the length of one bar (due to symmetry), because this problem involves only two degrees of freedom that are located at the same node. As this length of the trajectory is initially unknown, the integration limits of the functional are not fixed and remain unknown, as it was the case in the Brachistochrone problem. 

Therefore, another parameter must be introduced that indicates the motion progress with fixed integration bounds. In quasi-static structural analysis, often a normalized pseudo-time $t$ is used, which runs from $t=0$ to $t=1$. This idea is adopted here and the motion parameter is re-defined as a normalized arc length of the deformation path. To use this path parameter as integration variable, a substitution is necessary,
%Therefore, again a normalized arc length $\sb$, running from 0 to 1, is introduced 
\begin{align}
\int_0^{\su} (\ldots) \ds = \int_0^1 (\ldots) \frac{\ds}{\dsb} \dsb = \int_0^1 (\ldots) \su \dsb.
\end{align}
The mapping parameter
\begin{align}
\su  & := \frac{\ds}{\dsb} = \frac{\frac{1}{V} \int_\OM \sqrt{\Rd u_1^2 + \Rd u_2^2 + \Rd u_3^2}\dO}{\dsb} \\
& = \frac{1}{V} \int_\OM \sqrt{\left(\frac{\Rd u_1}{\dsb}\right)^2 + \left(\frac{\Rd u_2}{\dsb}\right)^2 + \left(\frac{\Rd u_3}{\dsb}\right)^2} \dO \\
& = \frac{1}{V} \int_\OM
\sqrt{u_{1,\sb}^2 + u_{2,\sb}^2 + u_{3,\sb}^2} \dO
\label{eq:arc_length}
\end{align}
is referred to as total arc length and the functional $J$ transforms to
\begin{align}
J &= \int_0^1 \int_\OM \frac 12 \BE^\RT \BC \BE \dO s_\Ru \dsb = \min.
\label{eq:functional_arclength}
\end{align}

%%%%%%%%%%%%%%%%%%%%%%%%%%%%%%%%%%%%%%%%%%%%%%%

\subsubsection{First variation of the functional}

The strain and the total arc length are functions of the unknown displacements. The variation is computed according to the chain rule and set equal to zero,
\begin{align}
\de J & = \int_0^1 \bigg[ \int_\OM \de \BE^\RT \BC \BE \dO s_\Ru + \int_\OM \frac 12 \BE^\RT \BC \BE \dO \de s_\Ru \bigg] \dsb = 0 .
\label{eq:variation}
\end{align}

\subsection{Spatial discretization and path discretization}

\subsubsection{General concept}
\label{sec:general_concept}

In order to solve the variational problem in equation~(\ref{eq:variation}), two discretizations are introduced. They divide both the spatial domain and the path into elements. Thus, a continuous problem is transferred into a discrete problem with a finite amount of degrees of freedom. Those degrees of freedom are located at the nodes forming the elements.

\subsubsection{Spatial discretization}
\label{sec:space_discr}

First, a standard spatial discretization of $\OM$ is introduced. The same continuity requirements apply as in a standard non-linear structural finite element analysis. The space is divided into $n_\Rele$ subdomains $\OM_e$, the finite elements, on which integration is performed,
\begin{align}
\int_\OM (\ldots) \dO &= \sum_{e=1}^{n_\Rele} \int_{\OM_e} (\ldots) \dO_e .
\label{eq:space_discr_1}
\end{align}
The unknown displacement field is approximated by shape functions, interpolating the unknowns between discrete values at $n_{\Rnd,\Rele}$ nodes per element
\begin{align}
\Bu (\BX,s) & \approx \Bu_h (\BX,s) = \sum_{k=1}^{n_{\Rnd,\Rele}} N_k (\BX) \Bd_k (s) = \BN (\BX) \Bd (s).
\label{eq:space_discr_2}
\end{align}
With the mapping of $\OM_e$ to a reference element with the natural coordinates $\Bxi$ by a Jacobian $\BJ_e = \fracpt{\BX_e}{\Bxi}$ the interpolation of the displacement field, as well as the reference and current geometry in an isoparametric concept, can be expressed as
\begin{align}
\Bu_h (\Bxi,s) = \sum_{k=1}^{n_{\Rnd,\Rele}} N_k (\Bxi) \Bd_k (s) = \BN (\Bxi) \Bd (s), \\
\BX_h (\Bxi,s) = \sum_{k=1}^{n_{\Rnd,\Rele}} N_k (\Bxi) \BX_k = \BN (\Bxi) \BX, \\
\Bx_h (\Bxi,s) = \sum_{k=1}^{n_{\Rnd,\Rele}} N_k (\Bxi) \Bx_k (s) = \BN (\Bxi) \Bx (s).
\label{eq:space_discr_3}
\end{align}
In a Bubnov-Galerkin approach, identical interpolation functions are used for the approximation of the variations
\begin{align}
\de \Bu(\Bxi,s) \approx \de \Bu_h (\Bxi,s) = \sum_{k=1}^{n_{\Rnd,\Rele}} N_k
(\Bxi) \de\Bd_k (s) = \BN (\Bxi) \de \Bd (s).
\end{align}
In order to obtain a system of equations for all parameters of the problem, the nodal values of the single elements need to be assembled. This can be formally written with an assembly operator
\begin{align}
\BD(s) = \bigcup_{e=1}^{n_\Rele} \Bd_e(s) ,
\end{align}
containing the information about element connectivity (topology). The nodal values of the displacement field and the current configuration still depend on the path variable $s$. The dimension of the vector $\BD(s)$ is equal to the number of spatial degrees of freedom $n_\Rdof$.

\subsubsection{Path discretization}
\label{sec:path_discr}

\begin{figure}[b]
\centering
\includegraphics[width=1.0\textwidth]{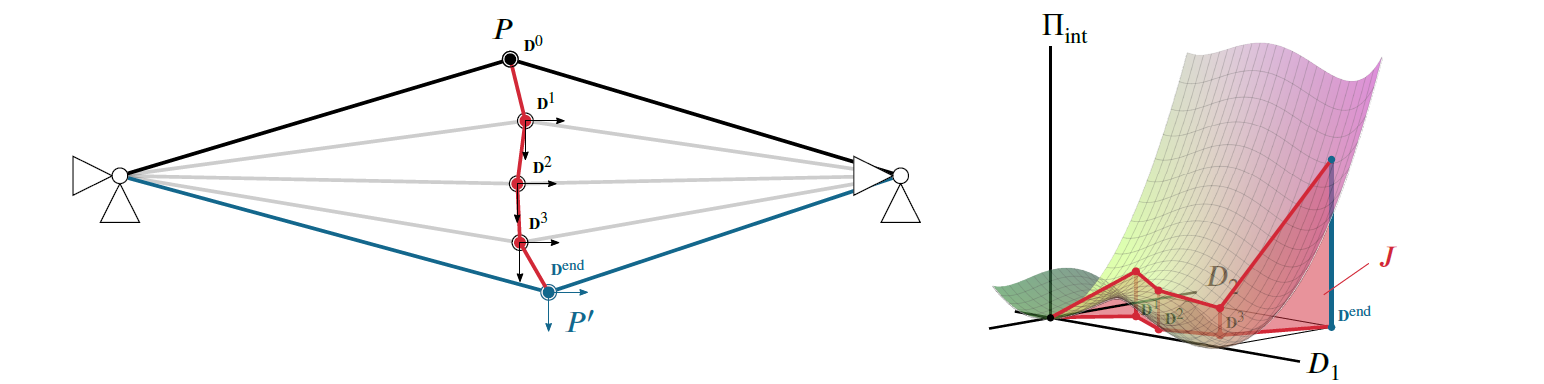}
\caption{Illustrating example with motion elements and the discretized illustration of the functional}
\label{fig:example_functional_discr}
\end{figure}

As the spatial degrees of freedom $\BD (s)$ are still functions of the path, a second discretization, the path discretization is required. It can also be denoted as a discretization of motion. This differs from a discretization in time, because the path also depends on the deformation of the structure, whereas time is considered an independent and autonomous value. The path, parametrized by the normalized arc length $\sb \in[0,1]$, is subdivided into $\bar{n}_{\Rele}$ path elements
\begin{align}
\int_0^1 (\ldots) \dsb = \sum_{{\bar{e}}=1}^{\bar{n}_{\Rele}} \int_{s^{\bar{e}}} (\ldots) \dsb^{\bar{e}} .
\end{align}
The shape functions can either be defined in the normalized parameter space $\sb \in [0,1]$ or they can be transformed by a Jacobian. Variables referring to the path discretization are marked with a bar $\bar{(\circ)}$. Element numbers are indicated with a superscript (instead of a subscript, as in spatial discretization) for distinction.
Also for the path elements, interpolation functions, a mapping to a reference element by a Jacobian $\bar{J}^{\bar{e}} = \fracpt{\sb^{\bar{e}}}{\bar{\xi}}$, as well as a Bubnov-Galerkin approach are used
\begin{align}
\Bd_{\bar{h}}(\bar{\xi}) & = \sum_{\bar{k}=1}^{\bar{n}_{\Rnd,\Rele}} \bar{N}^{\bar{k}} (\bar{\xi}) \bar{\Bd}^{\bar{k}} = \bar{\BN}(\Bxi) \bar{\Bd}, \\
\de \Bd_{\bar{h}}(\bar{\xi}) & = \sum_{\bar{k}=1}^{\bar{n}_{\Rnd,\Rele}} \bar{N}^{\bar{k}} (\bar{\xi}) \de \bar{\Bd}^{\bar{k}} = \bar{\BN}(\Bxi) \de \bar{\Bd}.
\end{align}
The nodes of the path discretization $\bar{k}$ represent the different geometric configurations throughout the motion, including the initial, intermediate and end configurations,
\begin{align}
\bar{\Bd}^{\bar{k}} & = \BD^{\bar{k}} = \BD(s=s^{\bar{k}}) \\
\de \bar{\Bd}^{\bar{k}} & = \de \BD^{\bar{k}} = \de \BD(s=s^{\bar{k}}) .
\end{align}
The shape functions serve for interpolation between the individual configurations and the total degrees of freedom are all $n_\Rdof$ spatial degrees of freedom in every configuration $\bar{k}$. Therefore, the vector $\bar{\Bd}$ consists of $\bar{n}_{\Rn\Rd,\Rele}$ subvectors
\begin{align}
\bar{\Bd} & = \begin{bmatrix} \BD^{1} & \BD^{2} & \ldots & \BD^{\bar{k}} & \ldots & \BD^{\bar{n}_{\Rn\Rd,\Rele}} \end{bmatrix}^\RT \\
\de \bar{\Bd} & = \begin{bmatrix} \de \BD^{1} & \de \BD^{2} & \ldots & \de \BD^{\bar{k}} & \ldots & \de \BD^{\bar{n}_{\Rn\Rd,\Rele}} \end{bmatrix}^\RT ,
\end{align}
where the length of $\bar{\Bd}$ is the amount of total degrees of freedom in one path element
\begin{align}
\bar n_\Rdof = \bar{n}_{\Rn\Rd,\Rele} \cdot n_\Rdof.
\end{align}
Path elements are one-dimensional. In the case of two spatial degrees of freedom, as in the illustrative example (Figure~\ref{fig:example_functional_discr}), the path elements discretize the trajectory of point $P$. With an increasing number of degrees of freedom in space, they form a one-dimensional subspace within an $n_\Rdof$-dimensional hyperspace.

The matrix of shape functions can contain any type of function. In Figure~\ref{fig:example_functional_discr} a discretization with linear Lagrange shape functions is illustrated, but B-splines and higher order functions are also possible. As the variational index in the calculation of the arc length is equal to 1, at least $C^{0}$-continuous functions are needed. Assembly is performed in the same manner as in spatial discretization with the assembly operator
\begin{align}
\bar{\BD} = \bigcup_{\bar{e}=1}^{\bar{n}_\Rele} \bar{\Bd}_{\bar{e}} .
\end{align}

Path discretization is further visualized in the illustrative example of the two bar truss in Figure~\ref{fig:example_functional_discr}, where the vectors $\BD^{\bar{k}}$ are displayed. In this example, the vectors consist of two components due to the two spatial degrees of freedom
\begin{align}
\BD^{\bar{k}} = \begin{bmatrix} D_1^{\bar{k}} & D_2^{\bar{k}} \end{bmatrix}^\RT .
\end{align}
For this prescribed path discretization by four linear elements, the degrees of freedom per path element are
\begin{align}
\bar{\Bd}_1&=\begin{bmatrix} D_1^0 & D_2^0 & D_1^1 & D_2^1 \end{bmatrix}^\RT \\
\bar{\Bd}_2&=\begin{bmatrix} D_1^1 & D_2^1 & D_1^2 & D_2^2 \end{bmatrix}^\RT \\
\bar{\Bd}_3&=\begin{bmatrix} D_1^2 & D_2^2 & D_1^3 & D_2^3 \end{bmatrix}^\RT \\
\bar{\Bd}_4&=\begin{bmatrix} D_1^3 & D_2^3 & D_1^\Rend & D_2^\Rend \end{bmatrix}^\RT
.
\end{align}
The vector with all degrees of freedom can then be built by assembly
\begin{align}
\bar{\BD}_1&=\begin{bmatrix} D_1^0 & D_2^0 & D_1^1 & D_2^1 & D_1^2 & D_2^2 & D_1^3 & D_2^3 & D_1^\Rend & D_2^\Rend \end{bmatrix}^\RT
,
\end{align}
where the parameters $D_1^0=0$, $D_2^0=0$, $D_1^\Rend$ and $D_2^\Rend$ are already defined in the problem formulation as the initial and the target configuration. The method for motion design therefore aims to move the intermediate configurations such that the functional $J$ is minimized. 

One issue with the motion design problem described so far is its potential ill-posedness for special cases. Within path discretization, nodes may be located anywhere on the trajectory, while still approximating the same curve (take a straight line as the simplest example for such a situation). Thus, the solution is no unique. This issue is well-known from, for instance, shape optimization and form finding problems of thin-walled structures, where nodes can be dislocated in-plane without changing the geometry. A corresponding regularization can be realized by either enforcing a constant path element size or by controlling the increments of a specified displacement degree of freedom throughout the deformation process. This aspect is further elaborated in Section~\ref{sec:convergence}.

At a first glance, path discretization resembles time integration in dynamic problems by space-time finite elements. However, both approaches are fundamentally different, since the arc length depends on the deformation of the structure, whereas time represents an independent and autonomous value. Another difference lies in the application of the two approaches. While space-time elements are mostly used to calculate and represent dynamic problems containing inertia effects, the motion path discretization is developed for quasi-static loading situations and static problems. This has an impact on the required element size, as dynamic effects, which can potentially be missed by using a time discretization that is too coarse, do not play a role in motion design problems.

In the next section, the discretized variations of the individual terms are introduced, where the two discretizations for space and path are introduced separately and successively.

%%%%%%%%%%%%%%%%%%%%%%%%%%%%%%%%%%%%%%%%%%%%%%%%%%%%%%%%%%%%%%%%%%%%%%%%%%%%%%%%

\subsection{Discretized variation and linearization}

\subsubsection{Spatial discretization}

\paragraph{Green-Lagrange strains}

For improved readability, in the following parameters depending on the path variable $s$ are written with a superscript $s$
\begin{align}
\Big(\bullet\Big)(s) =: \Big(\bullet\Big)^s.
\end{align}
The first variation of the Green-Lagrange strain with respect to the spatially discrete parameters $\Bd(s)$ can then be written as
\begin{align}
\de \BE^s & = \bigg(\fracpt{\BE^s}{\Bd^s}\bigg)^\RT \de \Bd^s .
\end{align}
The strain-displacement matrix
\begin{align}
\BB^s = \bigg(\fracpt{\BE^s}{\Bd^s}\bigg)^\RT
\end{align}
of the spatial elements is still continuous in the path variable $s$.

\paragraph{Total arc length}

\begin{figure}[h]
\centering
\includegraphics[width=1.0\textwidth]{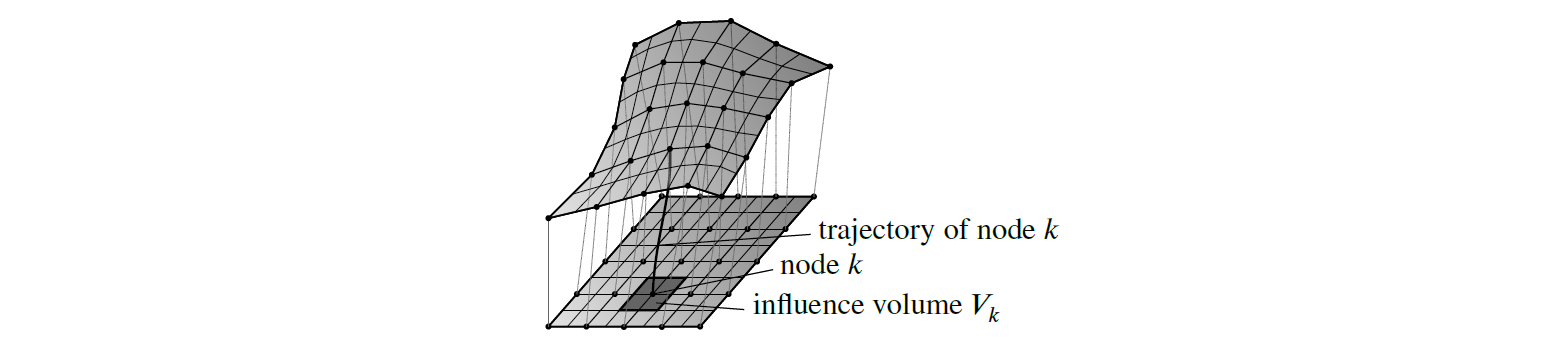}
\caption{Illustration of the spatially discretized total arc length}
\label{fig:arclength_discr}
\end{figure}

The total arc length from eq.~(\ref{eq:arc_length}) is now expressed in a spatially discretized form. First, the lengths of the trajectories of the individual nodes of the spatial discretization are generated.
The length of the nodal trajectories follows as
\begin{align}
s_{\Ru,k}^s = \sqrt{\sum_i^{n_{\text{disp,nd}}} D_{i,k}(s)^2 }. 
\end{align}
A simple summation of the lengths of the nodal trajectories results in a dependency of the spatial discretization. This would mean that the total arc length of a motion of a coarse spatial discretization is smaller compared to the one of a finer mesh. Therefore, a mean value, in this case the root mean square, of the nodal trajectory lengths is determined. To calculate the average value, the influence volume $V_k$ of each individual node $k$ is determined as illustrated in Figure~\ref{fig:arclength_discr}. The root mean square can then be computed as
\begin{align}
s_\Ru^s = \sqrt{ \frac{1}{V} \sum_{k}^{n_\text{nd}} V_k s_{\Ru,k}^2}.
\end{align}
This represents the spatially discretized total arc length.

As the total arc length $\su^s$ depends on the derivative of the total displacements only, its first variation is
\begin{align}
\de s_\Ru^s & = \bigg(\fracpt{s_\Ru^s}{\BD_{,s}^{s}}\bigg)^\RT \de\BD_{,s}^{s} ,
\end{align}
which includes the gradient of $\su$ with respect to the derivatives of the displacement degrees of freedom.

For a concise notation, the following abbreviations for the derivatives of the total arc length with respect to the spatial parameters are introduced
\begin{align}
\Bs_\Ru^s & := \fracpt{s_\Ru^s}{\BD_{,s}^s}, &
\BS_\Ru^s & := \frac{\pt^2s_\Ru^s}{(\pt\BD_{,s}^{s})^2}.
\end{align}

%%%%%%%%%%%%%%%%%%%%%%%%%%%%%%%%%%%%%%%%%%%%%%%%%%%%

\paragraph{Variation}

Together with eq.~(\ref{eq:variation}) the discretized variation of the functional reads
\begin{align}
\de J & = \int_0^1 \bigg[ \sum_{e=1}^{n_\Rele} \int_{\OM_e} (\de \Bd^s)^\RT \bigg(\fracpt{\BE^s}{\Bd^s}\bigg) \BC\BE^s \dO_e s_\Ru^s +
(\de\Bd_{,s}^s)^\RT \Bs_\Ru^s \sum_{e=1}^{n_\Rele} \int_{\OM_e} \frac 12 \BE^{s\RT} \BC\BE^s \dO_e \bigg] \dsb = 0 \,.
\label{eq:variation_discr}
\end{align}
As the vectors $\de \Bd^s$ and $\de \Bd_{,s}^s$ only contain discrete values in $\OM$, they can be extracted from the integral and by assembly eq.~(\ref{eq:variation_discr}) can be written as
\begin{align}
\de J & = \int_0^1 \bigg[ (\de \BD^s)^\RT \underbrace{\bigcup_{e=1}^{n_\Rele} \int_{\OM_e} (\BB^s)^\RT \BS^s \dO_e}_{\BF_\Rint^s} s_\Ru^s + (\de\BD_{,s}^s)^\RT \Bs_\Ru^s \underbrace{\bigcup_{e=1}^{n_\Rele} \int_{\OM_e} \frac 12 \BE^{s\RT} \BS^s \dO_e}_{\PI_\Rint^s} \bigg] \dsb = 0,
\end{align}
where $\bigcup$ denotes the usual assembly operator.
In this equation, the global vector of internal forces $\BF_\Rint^s$ and the internal energy $\PI_\Rint^s$, both still continuous in the path $s$, are identified
\begin{align}
\de J & = \int_0^1 \bigg[ (\de \BD^s)^\RT \BF_\Rint^s s_\Ru^s + (\de\BD_{,s}^s)^\RT \Bs_\Ru^s \PI_\Rint^s \bigg] \dsb = 0
\end{align}

\paragraph{Linearization}

Both terms can now be linearized separately
\begin{align}
\RLIN (\BF_\Rint^s s_\Ru^s) & = \BF_\Rint^s s_\Ru^s + \fracpt{\BF_\Rint^s s_\Ru^s}{\BD^s} \DE \BD^s + \fracpt{\BF_\Rint^s s_\Ru^s}{\BD_{,s}^s} \DE \BD_{,s}^s \\
& = \BF_\Rint^s s_\Ru^s + \bigg(\underbrace{\fracpt{\BF_\Rint^s}{\BD^s}}_{=\BK_\RT^s} s_\Ru^s +
\BF_\Rint^s \underbrace{\fracpt{s_\Ru^s}{\BD^s}}_{=0}\bigg) \DE \BD^s + \bigg(\underbrace{\fracpt{\BF_\Rint^{s}}{\BD_{,s}^s}}_{=0} s_\Ru^s + \BF_\Rint^s\underbrace{\fracpt{s_\Ru^s}{\BD_{,s}^s}}_{\Bs_\Ru^s}\bigg) \DE \BD_{,s}^s \\
& = \BF_\Rint^s s_\Ru^s + \BK_\RT^s s_\Ru^s \DE \BD^s + \BF_\Rint^s\Bs_\Ru^s \DE \BD_{,s}^s \\[5mm]
\end{align}
\begin{align}
\RLIN (\Bs_\Ru^s \PI_\Rint^s) & = \Bs_\Ru^s \PI_\Rint^s + \fracpt{\Bs_\Ru^s \PI_\Rint^s}{\BD^s} \DE \BD^s + \fracpt{\Bs_\Ru^s \PI_\Rint^s}{\BD_{,s}^s} \DE \BD_{,s}^s \\
& = \Bs_\Ru^s \PI_\Rint^s + \bigg(\underbrace{\fracpt{\Bs_\Ru^s}{\BD^s}}_{=0} \PI_\Rint^s + \Bs_\Ru^s \underbrace{\fracpt{\PI_\Rint^s}{\BD^s}}_{=\BF_\Rint^s} \bigg) \DE \BD^s + \bigg(\underbrace{\fracpt{\Bs_\Ru^s}{\BD_{,s}^s}}_{\BS_\Ru^s} \PI_\Rint^s + \Bs_\Ru^s \underbrace{\fracpt{\PI_\Rint^s}{\BD_{,s}^s}}_{=0} \bigg) \DE \BD_{,s}^s) \\
& = \Bs_\Ru^s \PI_\Rint^s + \Bs_\Ru^s \BF_\Rint^s \DE \BD^s + \BS_\Ru^s \PI_\Rint^s \DE \BD_{,s}^s.
\end{align}
Inserting these terms into the variation leads to the spatially discretized linearized variation
\begin{align}
\de J = \int_0^1 & \bigg[ (\de \BD^s)^\RT \Big(\BF_\Rint^s s_\Ru^s + \BK_\RT^s s_\Ru^s \DE \BD^s + \BF_\Rint^s\Bs_\Ru^s \DE \BD_{,s}^s \Big) \\
& + (\de\BD_{,s}^s)^\RT \Big( \Bs_\Ru^s \PI_\Rint^s + \Bs_\Ru^s \BF_\Rint^s \DE \BD^s + \BS_\Ru^s \PI_\Rint^s \DE \BD_{,s}^s \Big) \bigg] \dsb = 0.
\label{eq:variation_discr_lin}
\end{align}

\subsubsection{Path discretization}

Eq.~(\ref{eq:variation_discr_lin}) is still continuous along the path, so the path discretization from Section~\ref{sec:path_discr} is introduced for the parameters $\Bd^s$, their variation $\de\Bd^s$, the linearized parameters $\DE\Bd^s$ and the corresponding partial derivatives
\begin{align}
\BD^s & = \bar{\BN} \bar{\BD} &
\de\BD^s & = \bar{\BN} \de\bar{\BD} &
\DE\BD^s & = \bar{\BN} \DE\bar{\BD} \\
\BD_{,s}^s & = \bar{\BN}_{,s} \bar{\BD} &
\de\BD_{,s}^s & = \bar{\BN}_{,s} \de\bar{\BD} &
\DE\BD_{,s}^s & = \bar{\BN}_{,s} \DE\bar{\BD}.
\end{align}
Inserting the path discretization into eq.~(\ref{eq:variation_discr_lin}) yields the completely discretized and linearized variation
\begin{align}
\de J = \sum_{{\bar{e}}=1}^{\bar{n}_{\Rele}} \int_{s^{\bar{e}}} & \bigg[\de\bar{\BD}^\RT \bar{\BN}^\RT \Big(\BF_\Rint s_\Ru + \BK_\RT s_\Ru \bar{\BN} \DE\bar{\BD} + \BF_\Rint\Bs_\Ru \bar{\BN}_{,s} \DE\bar{\BD} \Big) \\
& + \de\bar{\BD}^\RT \bar{\BN}_{,s}^\RT \Big( \Bs_\Ru \PI_\Rint + \Bs_\Ru \BF_\Rint \bar{\BN} \DE\bar{\BD} + \BS_\Ru \PI_\Rint \bar{\BN}_{,s} \DE\bar{\BD} \Big) \bigg] \dsb^{\bar{e}} = 0.
\label{eq:variation_discr_lin2}
\end{align}
This can further be modified by extraction of $\de \bar{\BD}$ from the integral and by rearranging the terms to
\begin{align}
\de J & = \sum_{{\bar{e}}=1}^{\bar{n}_{\Rele}} \de \bar{\BD}^\RT  \bigg[ \int_{\sb^{\bar{e}}} \Big( \bar{\BN}^\RT \BF_\Rint s_\Ru + \bar{\BN}_{,s}^\RT \Bs_\Ru \PI_\Rint \Big) \ds^{\bar{e}} + \\ &
\phantom{= \de \bar{\BD}^\RT \sum_{{\bar{e}}=1}^{\bar{n}_{\Rele}} \bigg[ }
\int_{s^{\bar{e}}} \Big( \bar{\BN}^\RT \BK_\RT s_\Ru \bar{\BN} + \bar{\BN}^\RT \BF_\Rint\Bs_\Ru \bar{\BN}_{,s} + \bar{\BN}_{,s}^\RT \Bs_\Ru \BF_\Rint \bar{\BN} + \bar{\BN}_{,s}^\RT \BS_\Ru \PI_\Rint \bar{\BN}_{,s} \big) \ds \DE \bar{\BD} \bigg] \dsb^{\bar{e}} = 0
\end{align}

\subsection{Global linearized system of equations}
\label{sec:system_of_equation}

From the condition that the discretized variation must vanish for any $\de\bar{\Bd}$ the following system of equations can be derived
\begin{align}
\bigcup_{{\bar{e}}=1}^{\bar{n}_{\Rele}} \int_{\sb^{\bar{e}}} \Big( \bar{\BN}^\RT \BK_\RT s_\Ru \bar{\BN} + \bar{\BN}^\RT \BF_\Rint\Bs_\Ru \bar{\BN}_{,s} + \bar{\BN}_{,s}^\RT \Bs_\Ru \BF_\Rint \bar{\BN} + \bar{\BN}_{,s}^\RT \BS_\Ru \PI_\Rint
\bar{\BN}_{,s} \Big) \ds^{\bar{e}} \DE \bar{\BD} \\
 = - \bigcup_{{\bar{e}}=1}^{\bar{n}_{\Rele}} \int_{s^{\bar{e}}} \Big( \bar{\BN}^\RT \BF_\Rint s_\Ru + \bar{\BN}_{,s}^\RT \Bs_\Ru
\PI_\Rint \Big) \dsb^{\bar{e}}.
\end{align}
With the definitions
\begin{align}
\BK_\Rmod & = \bigcup_{{\bar{e}}=1}^{\bar{n}_{\Rele}} \int_{\sb^{\bar{e}}} \Big( \bar{\BN}^\RT \BK_\RT s_\Ru \bar{\BN} + \bar{\BN}^\RT \BF_\Rint\Bs_\Ru \bar{\BN}_{,s} + \bar{\BN}_{,s}^\RT \Bs_\Ru \BF_\Rint \bar{\BN} + \bar{\BN}_{,s}^\RT \BS_\Ru \PI_\Rint \bar{\BN}_{,s} \Big) \dsb^{\bar{e}} \\
\BR_\Rmod & = \bigcup_{{\bar{e}}=1}^{\bar{n}_{\Rele}} \int_{\sb^{\bar{e}}} \Big( \bar{\BN}^\RT \BF_\Rint s_\Ru +
\bar{\BN}_{,s}^\RT \Bs_\Ru \PI_\Rint \Big) \dsb^{\bar{e}}
\end{align}
we obtain the system of equations in the familiar format,
\begin{align}
\BK_{\Rm\Ro\Rd} \DE \bar{\BD} & = -\BR_{\Rm\Ro\Rd}.
\label{eq:sys_eq}
\end{align}
Note that with this system the entire problem is solved monolithically, instead of incrementally proceeding along the path. On convergence of the iterative solution method, all intermediate configurations along the path are obtained in one go.

The system of equations depends on the used element as it includes the stiffness matrix and the internal forces. However, all ingredients can be combined in a modular manner. Thus, it does not pose any problem to use various element types, like mixed elements or isogeometric spatial discretizations. 

\begin{figure}[b]
\centering
\includegraphics[width=1.0\textwidth]{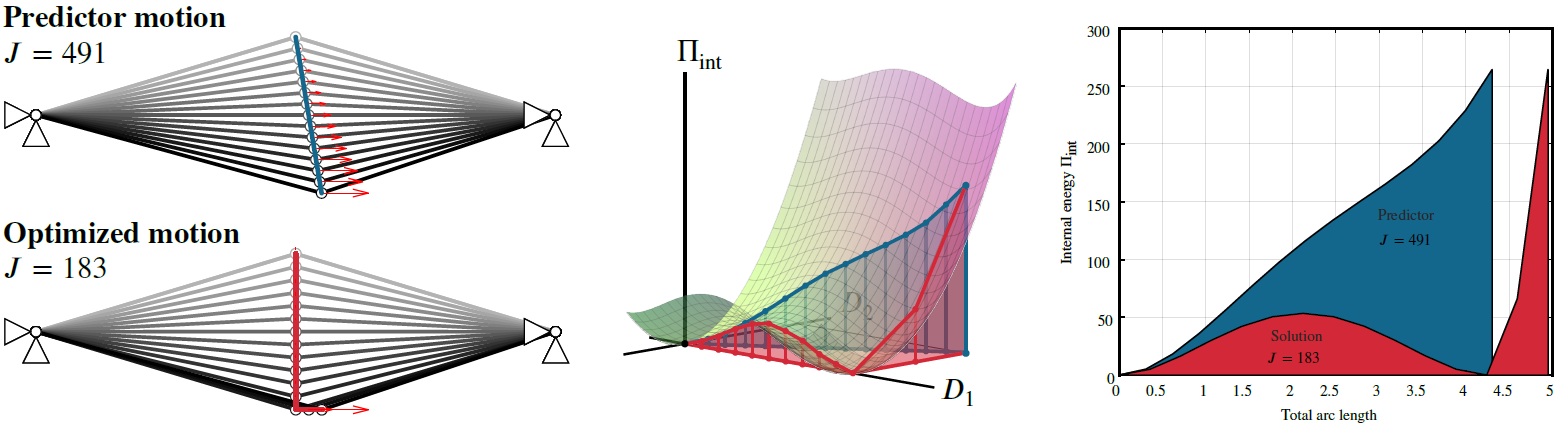}
\caption{Predictor motion and solution of illustrating example}
\label{fig:example_solution}
\end{figure}

In the solution of the illustrating two bar truss problem, the path is discretized by 14 linear path elements. The predictor represents an entire motion and it is chosen as a linear interpolation between the initial configuration and the target configuration.

The solution process of the non-linear problem with Newton's method converges after 9~iterations below the tolerance value of $10^{-8}$ of the $L_2$ norm of the residual. As a result of this simple motion design problem, it is found, that for minimizing the integrated internal energy it is beneficial to first enforce a purely vertical snap-through, followed by a horizontal movement (as opposed to following the straight path of the linear predictor motion). Therefore, the motion design method yields an optimized motion in a purely formalized way without the need to put any engineering expert knowledge into the analysis. The trajectory of the midpoint (as well as the path) is longer in the solution than in the predictor, but the proposed detour leads to a smaller accumulated internal energy throughout the motion. The difference is visualized in a plot of the internal energy over the two spatial degrees of freedom in Figure~\ref{fig:example_solution} (center) and a projection of the resulting functional surfaces where the internal energy is plotted versus the arc length in Figure~\ref{fig:example_solution} (right). The snap-through characteristics can also be detected in the progress of the internal energy for the final solution. After snap-through, the internal energy reaches the value zero.

In order to realize the prescribed deformation that results from motion design, forces are needed. Those forces are evaluated after convergence from the internal forces and equilibrium of internal and external forces. This means that for the special case considered so far, where all degrees of freedom are controlled, the equilibrium conditions are not needed for the solution of the motion design problem, but the equilibrium equations can be used in post-processing of the nodal forces.

\subsection{A generalized system of equations for any objective function}

So far, the minimization of the internal energy, integrated along the path, was used as a proof of concept for the proposed motion design framework. However, in principle any functional or objective function can be used. In general, we define a quantity $F$ that depends on the displacements and integrate it along the path to obtain the generalized functional 
\begin{align}
J &= \int_0^1 F s_\Ru \dsb = \min.
\end{align}
After spatial discretization, the variation is built by the product rule
\begin{align}
\de J & = \int_0^1 \Big[ \de (\BD^s)^\RT F^s_{,\BD^s} s_\Ru + (\de\BD_{,s}^s)^\RT \Bs_\Ru^s F^s \Big] \dsb  = 0 .
\end{align}
The linearization can then be derived for the two terms
\begin{align}
\RLIN & \big(F^s_{,\BD^s} s_\Ru\big) = F^s_{,\BD^s} s_\Ru + \big(F^s_{,\BD^s\BD^s} s_\Ru^s \big) \DE \BD^s + \big(F^s_{,\BD^s} \Bs_\Ru^s\big) \DE \BD_{,s}^s \\[5mm]
\RLIN & \big(\Bs_\Ru^s F^s\big) = \big(\Bs_\Ru^s F^s\big) + \big( \Bs_\Ru^s F^s_{,\BD^s} \big) \DE \BD^s + \big(\BS_\Ru^s F^s \big) \DE \BD_{,s}^s.
\end{align}
Path discretization and reordering the terms leads to 
\begin{align}
\sum_{{\bar{e}}=1}^{\bar{n}_{\Rele}} \int_{\sb^{\bar{e}}} \Big( \bar{\BN} F_{,\BD^s\BD^s} s_\Ru \bar{\BN}
 + \bar{\BN} F_{,\BD^s} \Bs_\Ru \bar{\BN}_{,s}
 + \bar{\BN}_{,s}^\RT \Bs_\Ru F_{,\BD^s} \bar{\BN} + \bar{\BN}_{,s}^\RT \BS_\Ru F \bar{\BN}_{,s} \Big) \ds^{\bar{e}} \DE \bar{\BD} \\
 = - \sum_{{\bar{e}}=1}^{\bar{n}_{\Rele}} \int_{s^{\bar{e}}} \Big(\bar{\BN}^\RT F_{,\BD^s} s_\Ru + \bar{\BN}_{,s}^\RT \Bs_\Ru F \Big) \dsb^{\bar{e}}.
\end{align}
This is the system of equations for a general objective function for which analytical derivatives can be calculated. The minimized quantity must depend on the displacements to apply this method for motion design, but they can as well be calculated numerically. In a lot of cases, quantities and their derivatives, e.g. strains and stresses, can be used that are routinely available in standard finite element codes.

\subsection{Aspects of convergence}
\label{sec:convergence}

The derived non-linear problem needs to be solved iteratively. The degrees of freedom are all spatial degrees of freedom in every single configuration. Some configurations are known, like the initial, starting geometry and the final, target geometry. It is also possible to define only parts of the target geometry. As the entire motion has to be found by one monolithic solution of the system of equations, the predictor describes an entire motion. It can be seen in Figures~\ref{fig:example_functional},\ref{fig:example_functional_discr} and \ref{fig:example_solution} that is a problem with two spatial degrees of freedom in one point $P$, the path elements discretize the trajectory of this point $P$ until it reaches the endpoint $P'$. However, the distribution and length of the path elements are not yet specified, which leads to an ill-posed problem. This can be fixed by additional controls. Either the progression of one (or multiple) spatial degrees of freedom can be prescribed by e.g. constant increments between the configurations or equal length of the path elements can be enforced by the introduction of Lagrange multipliers and the corresponding changes in the system of equations. The second method, the use of Lagrange multipliers, was applied in the illustrating example (see Figure~\ref{fig:example_solution}).

As the difference between the predictor motion and the final result may be large, the solution process sometimes suffers from convergence problems and the Newton method occasionally diverges after a number of iterations. There is no straightforward analogy to incremental-iterative solution procedures with the option to decrease the size of the increments in order to improve convergence behavior. There are, however, some other measures that can be taken:

\begin{description}
\item[Fewer degrees of freedom by path approximation with B-splines] The path may be approximated more efficiently by B-spline functions compared to linear Lagrange functions. The resulting reduction in the number of degrees of freedom leads to better convergence behavior.

\item[Improved predictor from solution with coarse path discretization] Likewise, a calculation with a small number of path elements, and therefore fewer degrees of freedom, improves the convergence of the Newton algorithm. This might result in a poor approximation due to the coarse discretization. This solution, however, can be used as an improved predictor for a computation with a finer path discretization. More generally, a hierarchically modified predictor improves convergence.

\item[Better predictor by a preanalysis] This method also focuses on the improvement of the first guess, the predictor. Instead of a linear interpolation, a standard non-linear analysis of the structure can be carried out to already approach a feasible motion. To this end, the internal forces in the end configurations are taken as the external forces (load case) for the non-linear analysis. The obtained equilibrium path is then used as a predictor motion.

\item[Modification of the Newton method with a relaxation factor] To improve convergence, a modification of the Newton method can be applied in which a relaxation factor prevents off-shooting from a possible solution during iterations in which the norm of the residual increases. 
This method has been presented in~\cite{albanese_numerical_1992} and further investigated and developed in~\cite{fujiwara_method_1993}.
\end{description}

All the described methods can also be combined.

%%%%%%%%%%%%%%%%%%%%%%%%%%%%%%%%%%%%%%%%%%%%%%%%%%%%%%%%%%%%%

\section{Numerical experiments}\label{sec:num_exp}

Numerical experiments are presented to test and verify the potential of the proposed method for motion design. Some examples serve for quantitative benchmarking of the solution and others contribute to a better understanding of the solution and possible applications.

\subsection{Kinematic structures for benchmarking}

First, benchmarking examples, for which the exact solutions are known, are used for verification. Obvious scenarios are kinematic mechanisms for which the internal energy is identically zero throughout the entire motion. 

\subsubsection{Kinematic truss system}
\label{sec:bsp_kin_truss}

\begin{figure}[t]
\centering\small
\includegraphics[width=1.0\textwidth]{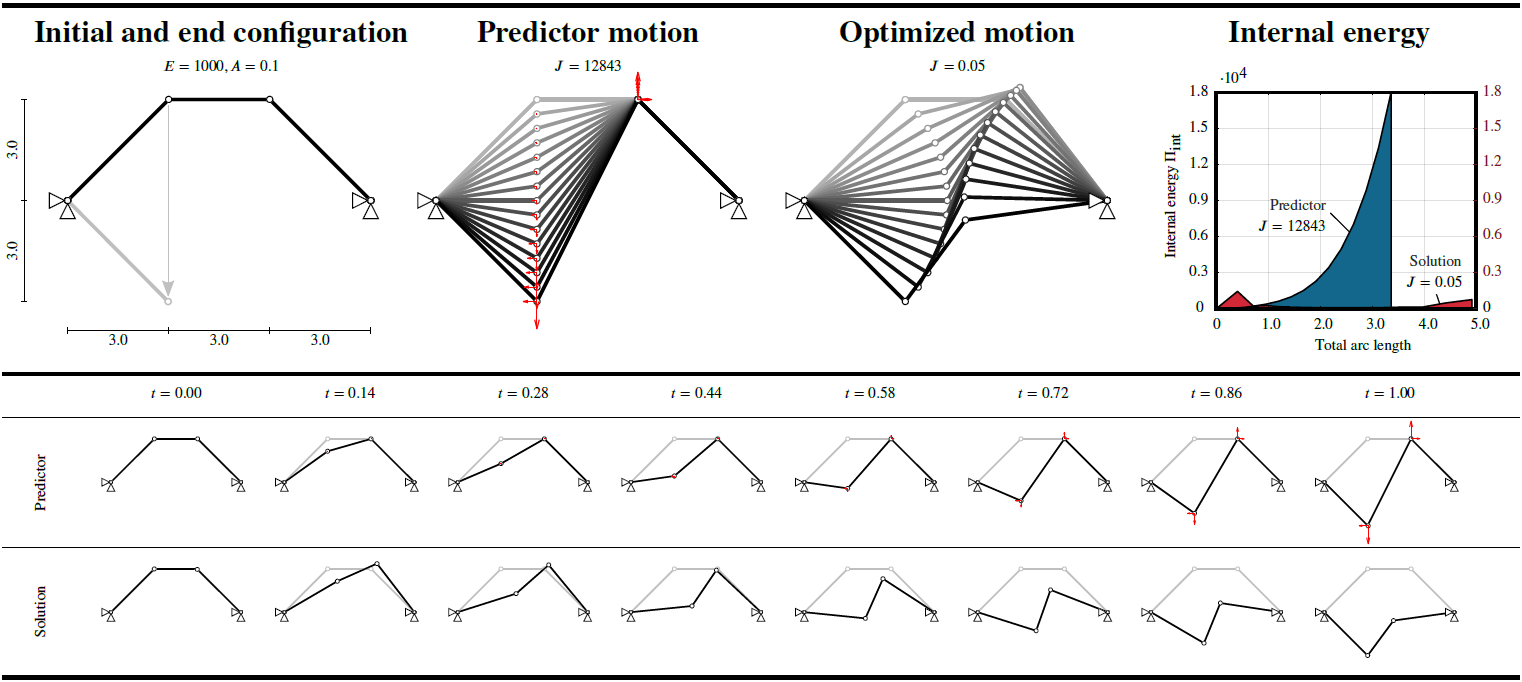}
\caption{Kinematic structure for benchmarking}
\label{fig:bsp_1}
\end{figure}

The first example is a kinematic truss system with four nodes, three bars and two supports, as shown in Figure~\ref{fig:bsp_1}. This kinematic system allows for a purely energy-free rigid body movement during which the lengths of the bars do not change.
Pretending that the target configuration is unknown, it is sufficient to specify the vertical displacement of the second node to obtain a well-posed problem. The other displacements are expected to adjust to allow the kinematic movement (to minimize the functional of motion design).

The path is discretized by 14 linear Lagrange elements, resulting in $\bar{n}_\Rdof=42$ degrees of freedom. For regularization, the vertical displacement of the second node is prescribed throughout the motion. The predictor is -- intentionally naive -- chosen to be a linear interpolation between the initial and the prescribed end configuration for the upper left node, while the upper right node does not move at all, as seen in Figure~\ref{fig:bsp_1}. As this is not a rigid body motion, forces are needed to enforce it, which are shown as red arrows. The predictor is far off the expected solution, with a functional value of $J = 12843$.

The table in Figure~\ref{fig:bsp_1} shows a comparison of seven snapshots, i.e. every second intermediate configuration, of the converged solution and the predictor motion. It can be seen that the solution obtained from motion design provides the expected result with zero length changes of the individual bars despite the naive predictor.
The difference between the linear interpolation and the final solution is also visible in the diagram on the right, where the internal energy is plotted versus the arc length along the path. The value of the functional represents the integral of the curve, i.e. the area of the blue and red area, respectively. The length of the path differs between the two motions. In the predictor motion, the moving node moves directly to the end position while the other node does not move at all. This results in a shorter path compared to the resulting path of the solution. The internal energy is much smaller as almost no strain is present in the bars. The fact that the energy is not exactly zero results from the error due to path discretization errors (note the difference in the y-axis of the factor $10^4$).

\begin{figure}[h]
\centering
\includegraphics[width=1.0\textwidth]{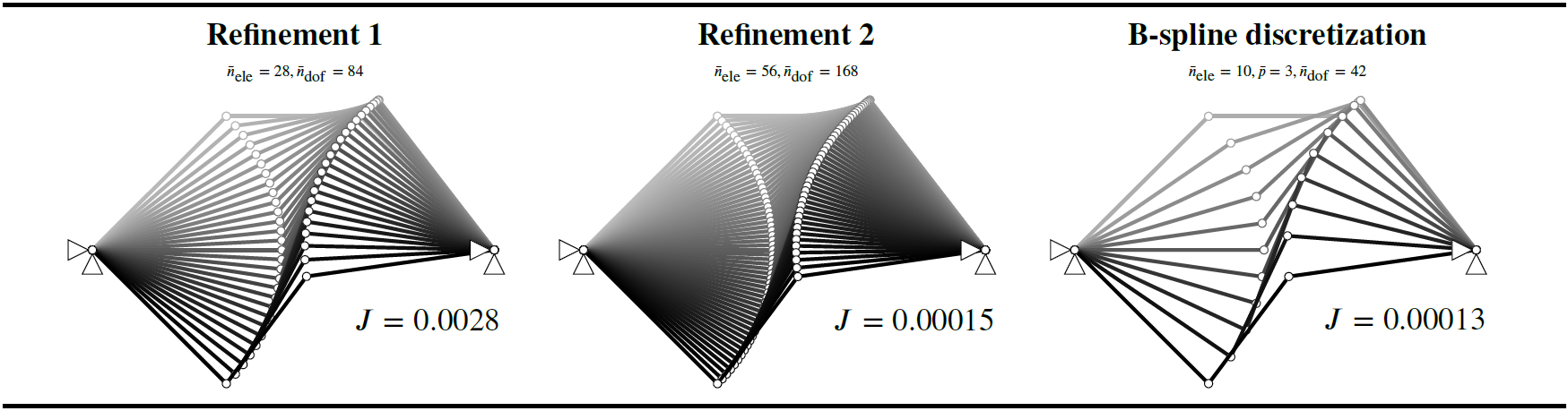}
\caption{Convergence study for kinematic structure}
\label{fig:bsp_1_konvergenz}
\end{figure}

The value of $J=0.05$ of the functional is not exactly zero for the obtained solution due to the error from path discretization with linear elements.
By a refinement of the path discretization, as seen in Figure~\ref{fig:bsp_1_konvergenz}, center, the approximation quality increases and the value of the functional approaches zero. The analysis with a discretization by B-splines, shown on the right, enables an even better approximation of the curved motion trajectory and results in a smaller value of the functional with fewer degrees of freedom.

\subsubsection{Folding motion with quadrilateral elements}

\begin{figure}[h]
\centering\small
\includegraphics[width=1.0\textwidth]{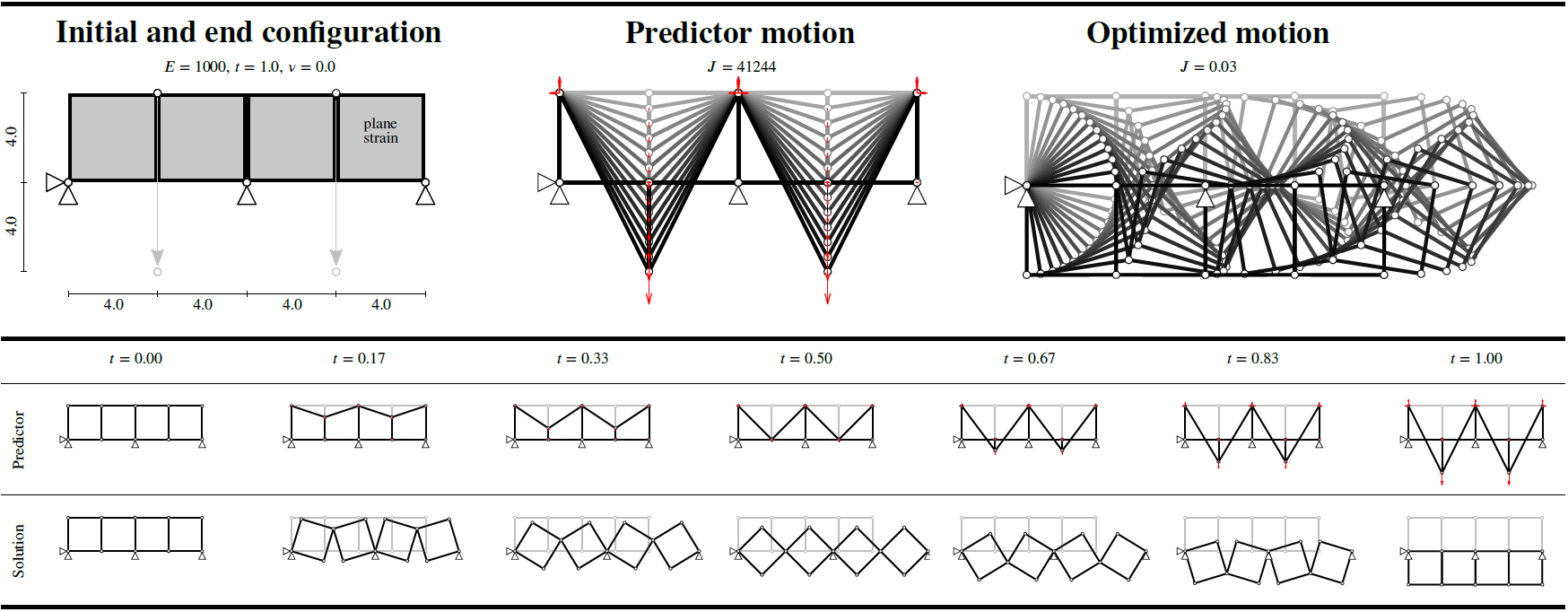}
\caption{Kinematic structure with quadrilateral elements}
\label{fig:bsp_2}
\end{figure}

In the next example, a fold-like motion of an assembly of four quadrilateral elements, as shown in Figure~\ref{fig:bsp_2}, is modeled. 
The elements are connected with hinges either on the upper or on the lower corner. This enables mirroring of the geometry solely by rigid body translations and rotations. The path is approximated with quadratic B-splines and $\bar{n}_\Rele=6$ elements. Again, the target geometry is assumed to be unknown, only the vertical displacement of the upper second and fourth node is prescribed and the vertical displacement increments are controlled during motion. The predictor motion is a linear interpolation and shows an unphysical movement with self-penetration of the elements.

By an iterative solution with the linearized system of equations presented in eq.~(\ref{eq:sys_eq}), the correct motion with zero internal energy throughout the motion is found (see Figure~\ref{fig:bsp_2}).

The analysis of the two kinematic structures verifies that the proposed method finds the correct solution for this specific class of problems.

\subsection{Motion design for problems with instabilities}

\subsubsection{Motivation}

Another interesting aspect to further understand and validate the properties of the proposed motion design method is the analysis of structures and motions where snap-through or bifurcation can occur. Various structures with potential instabilities are investigated next. A two-bar truss system that performs snap-through has already been presented during the derivation of the method in Chapter~\ref{sec:motion_design}.

\subsubsection{Motion design with multiple snap-through processes}

\begin{figure}[b]
\centering\small
\includegraphics[width=1.0\textwidth]{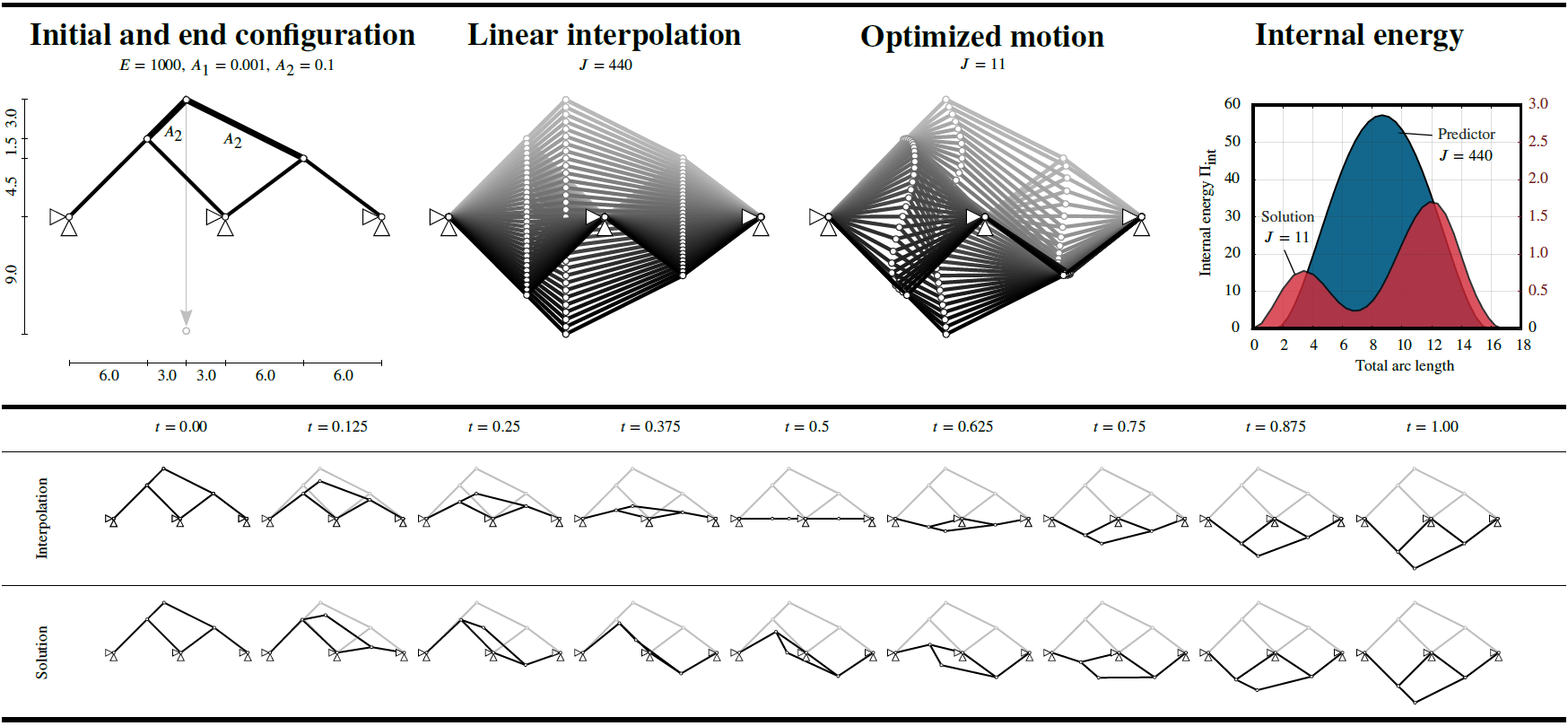}
\caption{Motion design with a combination of multiple snap-through}
\label{fig:bsp_3}
\end{figure}

The combination of three pairs of hinged bars, shown in Figure~\ref{fig:bsp_3}, represents a system for which the equilibrium path may exhibit multiple limit points, i.e. horizontal tangents, where snap-through occurs. The upper two-bar truss with a larger cross-sectional area $A_2$ is supported by two other two-bar trusses (cross section $A_1$) that can perform snap-through as well. The path is discretized by 32 linear elements. Only the vertical displacement of the upper node is prescribed and controlled throughout the motion. To enhance convergence behavior, the predictor is calculated and updated hierarchically from a solution obtained with a course path discretization as explained in Section~\ref{sec:convergence}.
Therefore the linear interpolation with 32 elements does not represent the predictor motion in this case. 

The solution is compared to a linear interpolation between the initial and a mirrored geometry, which is expected to represent a better approximation than the naive linear interpolation of only the upper central node to the target position. This results in a motion dominated by global snap-through. The result of motion design provides a different type of motion. When the side structures don't perform the snap-through at the same time, internal energy can be "saved" in the upper truss. This behavior can also be detected in the progress of the internal energy. The surface of the two ``snap-through bulges'' are clearly identifiable. The resulting end configuration is found to be the horizontally mirrored geometry, which reduces the internal energy back to zero. The value of the functional decreases significantly from $J=440$ to $J=11$.

\subsubsection{Motion design in a bifurcation problem}

\begin{figure}[h]
\centering\small
\includegraphics[width=1.0\textwidth]{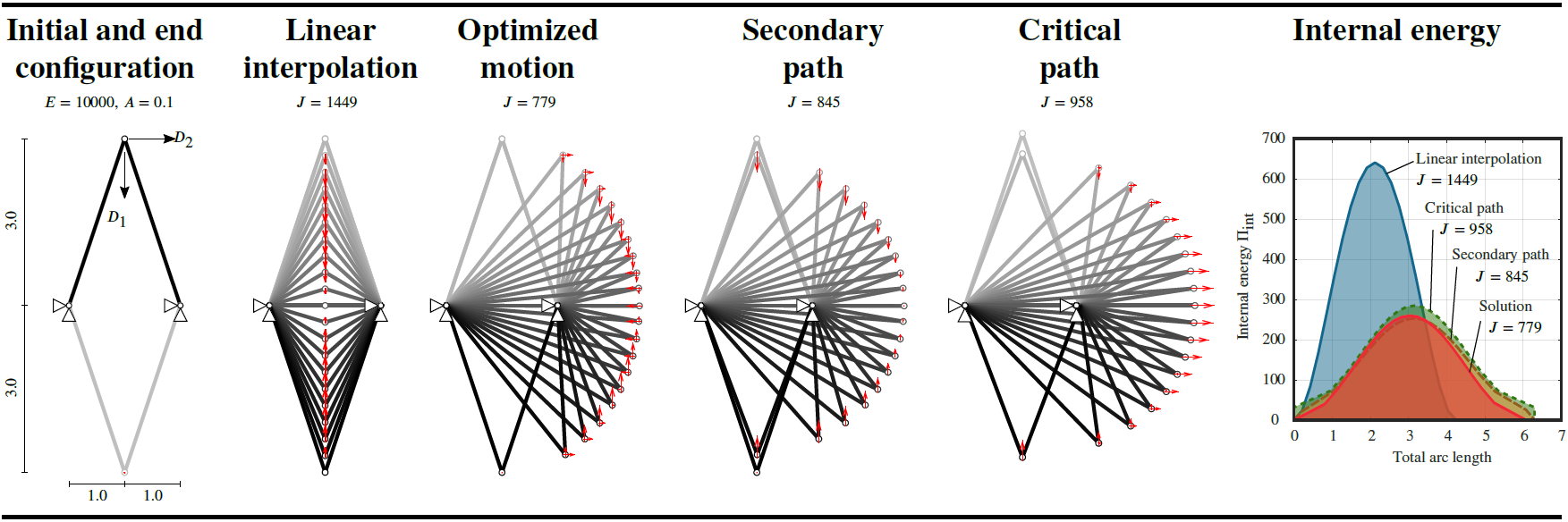}
\caption{Analysis of a two-bar truss with bifurcation and motion design}
\label{fig:bsp_bif}
\end{figure}

In a high two-bar truss subject to a vertical load, as shown in Figure~\ref{fig:bsp_bif}, bifurcation occurs before a limit point (snap-through) is reached, as it is the case in a shallow two-bar truss. Here, a system with a width-to-height ratio of $1:3$ is investigated. The path is discretized by 20 linear elements and the vertical displacement is controlled for motion design. For the vertically flipped geometry as target configuration, linear interpolation describes a purely vertical snap-through motion. Indeed, this happens to represent a stationary point for the functional of motion design. However, it provides a relative maximum of $J$, not a minimum, meaning that it is a worst case scenario. Therefore, the predictor needs to be modified significantly to improve convergence of the motion design algorithm to the desired solution.

For example, instead of a linear interpolation, a combination of the primary path -- up to the critical point -- followed by an arbitrarily chosen branch of the secondary equilibrium path, describing the deformation after buckling of the structure, can be used as predictor. The optimized motion found on the basis of this predictor is shown in Figure~\ref{fig:bsp_bif} and yields a functional value of $J=779$. It is significantly smaller than the value of $J=1449$ obtained from linear interpolation, the ``worst case scenario'' mentioned above. But it is also superior to the value $J=845$ obtained for the improved predictor based on the secondary path, which confirms the virtue of the method of motion design.

It can be observed, however, that the maximum value of the internal energy during deformation is higher for the optimized motion than for the secondary path (diagram on the right in Figure~\ref{fig:bsp_bif}). The fact that the functional value is still lower for the optimized motion follows from two aspects: During the first phase of the deformation process, the internal energy value is higher in the predictor than in the optimized motion and the deformation path is slightly longer. These aspects are dominant and lead to the reduction of the functional value, even though the maximum value of internal energy is higher in the optimized solution.

Yet an alternative predictor is the so-called \emph{critical path}. It is defined as the path that connects configurations for which the determinant of the stiffness matrix is zero, $\det \BK =0$. It leads to a functional value of $J=956$, which is worse than both the optimal solution and the solution obtained from following the secondary path. Nevertheless, it is a valid predictor for obtaining convergence of the motion design algorithm.

\subsubsection{Snap-through of a shallow arch}

\begin{figure}[t]
\centering\small
\includegraphics[width=1.0\textwidth]{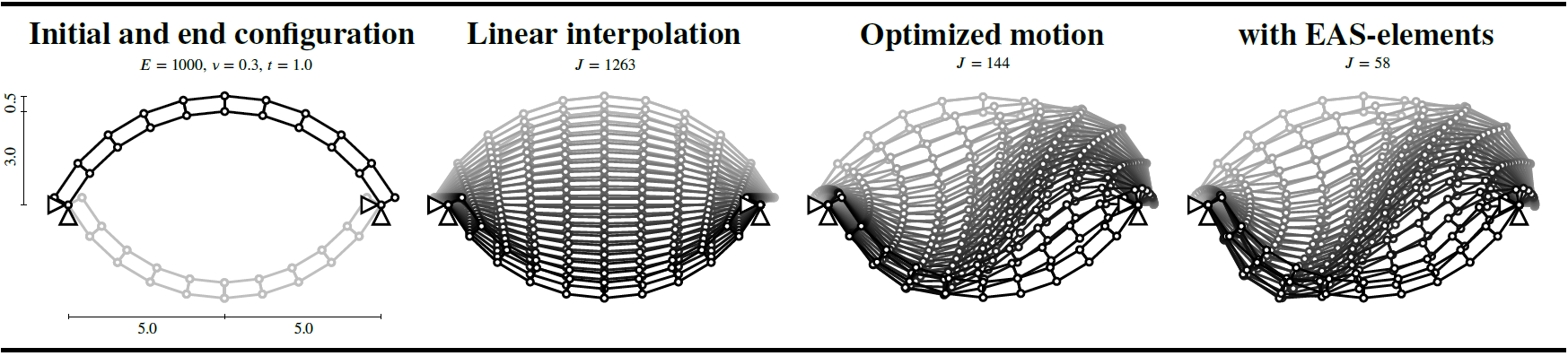}
\caption{Arch with quadrilateral elements and the influence of locking on motion design}
\label{fig:bsp_arc_Q1}
\end{figure}

The last example with a snap-through is a shallow arc, which is modeled as a two-dimensional structure under plane stress conditions, using quadrilateral finite elements, as shown in Figure~\ref{fig:bsp_arc_Q1}. The fully prescribed target geometry is artificially chosen and represents the (approximately) mirrored geometry of the initial configuration. The path is discretized by five elements with cubic B-splines as shape functions and the vertical displacement of the center node is controlled for motion design. 

First, purely displacement-based bilinear quadrilateral elements are used for spatial discretization. The predictor motion is again a linear interpolation between the initial and end configuration and represents a symmetric snap-through-dominated motion. By motion design, an antisymmetric swaying motion is found, which decreases the value of the functional from $J=1263$ to $J=144$. In this symmetric example, the mirrored deformation is equivalent to the calculated solution.

It is well known that displacement-based finite elements suffer from locking. Therefore, the influence of locking on the result of motion design is investigated next by using quadrilateral finite elements including an Enhanced Assumed Strain (EAS) formulation, proposed by~\cite{simo_class_1990}. The resulting stiffness matrix and internal forces of this formulation can simply be plugged into the system of equations from Chapter~\ref{sec:motion_design}. With four additional strain parameters per element, shear locking and volumetric locking can be eliminated. Already in the snapshots of the motion shown in Figure~\ref{fig:bsp_arc_Q1} the difference in the result obtained with finite elements that suffer from locking and locking-free elements is visible, although the overall character of the motion seems to be similar. Even though locking is not very dominant in this example, it can be observed that the EAS-elements exhibit more bending throughout the motion. The artificial energy that results from locking effects increases the internal energy along the path and acts as a penalty for bending modes.
Locking-free elements avoid this penalty and the value of the functional decreases significantly from $J=144$ to $J=58$.
It can therefore be expected that for structures that are more prone to locking, like slender thin-walled structures, locking has a significant effect on the result of motion design. Corresponding observations have been made for optimization problems in~\cite{camprubi_shape_2004}.

\subsection{Specification of intermediate configurations}

\begin{figure}[t]
\centering\small
\includegraphics[width=1.0\textwidth]{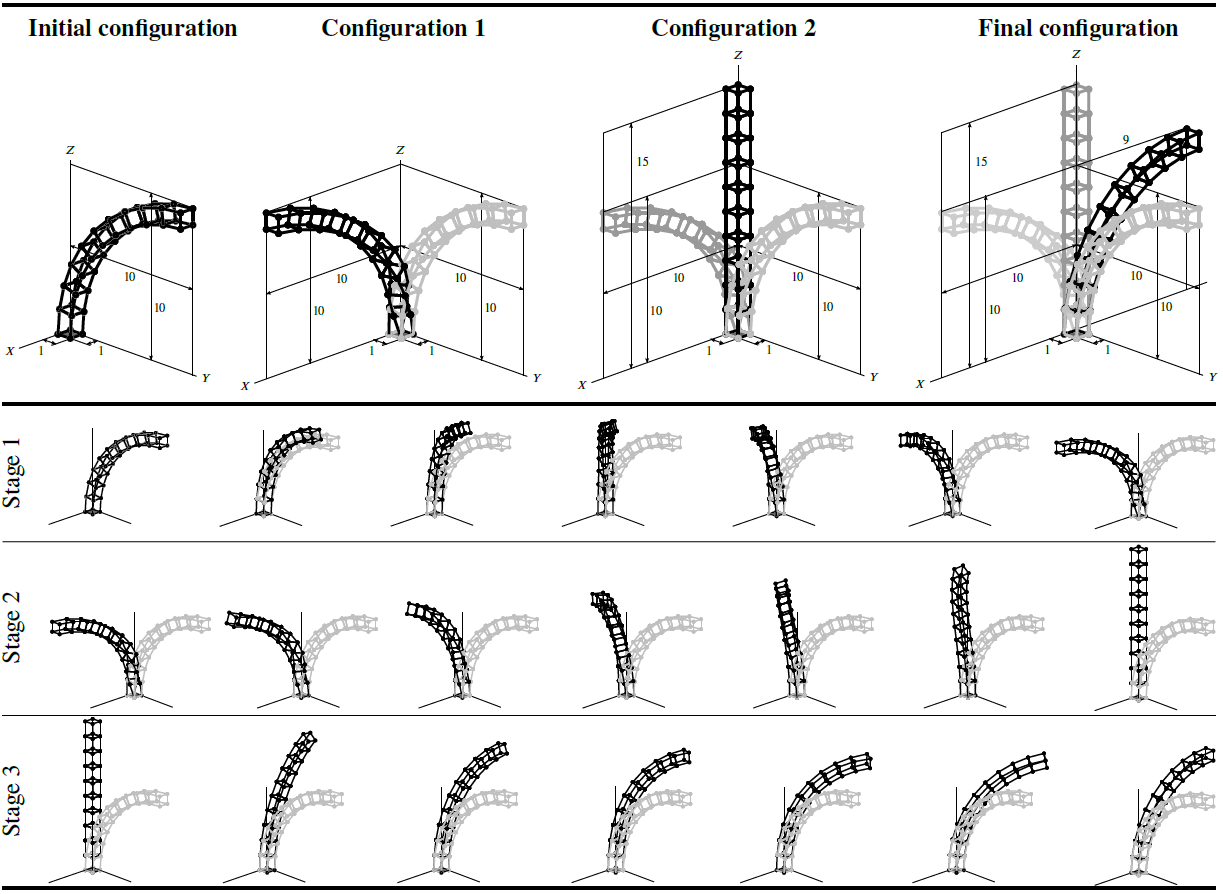}
\caption{Specification of intermediate configurations on a cantilever beam modelled with volume elements}
\label{fig:bsp_vol_cantilever}
\end{figure}

Beyond the possibility to specify initial and target configuration, also intermediate configurations can be included as an objective for motion design. Figure~\ref{fig:bsp_vol_cantilever} shows a three-dimensional curved cantilever beam, discretized by trilinear volume elements. There are two intermediate and a final target configuration. First, the cantilever tip is rotated by -90$^\circ$ around the $z$-axis (Configuration~1). Configuration~2 is defined as a straight, vertical bar. The final target configuration is identical to the initial configuration, rotated by 90$^\circ$ about the $z$-axis, as shown in Figure~\ref{fig:bsp_vol_cantilever}, right.

Path discretization is accomplished with a total of nine quadratic elements -- three elements for every deformation stage -- using B-splines as shape functions.
While path discretization with quadratic B-splines is usually $C^1$-continuous, continuity is reduced to $C^0$ at path nodes that correspond to the intermediate configurations in order to respect the expected non-smoothness of the solution.

For stabilization, the displacement in $y$-direction of one node at the cantilever tip was controlled in stage~1. In this case, the $C^1$-continuity of the path discretization was reduced to a $C^0$-continuity at the node of configuration~1.

The final solution with the individual stages is shown in Figure~\ref{fig:bsp_vol_cantilever}.

\subsection{Calculation of inextensible deformations of shells}

\subsubsection{Basic concept}

Motion design can also be performed for shells. One interesting option in this context is a modification of the functional by replacing the complete internal energy by the membrane energy only. This provides a method to compute motions that try to avoid membrane strains during deformation while bending remains without any penalization. The results are (nearly) inextensional deformations. Inextensional deformations of surfaces are defined as deformations that preserve lengths and angles of infinitesimal line elements at each point. Gaussian curvature remains constant during inextensional deformations. For thin shells (and beams) inextensional deformations can also be classified as pure bending deformations.

In the following examples, isogeometric Kirchhoff-Love elements, as presented in~\cite{kiendl_isogeometric_2009}, are used. It has to be noted that these elements still suffer from membrane locking, although by integration of the internal energy, strain oscillations are leveled out to a certain extent. However, when recovering the forces required to realize the found deformations, the effect of locking leads to values that are too large.

\subsubsection{Deformation of a cantilever beam}

\begin{figure}[b]
\centering\small
\includegraphics[width=1.0\textwidth]{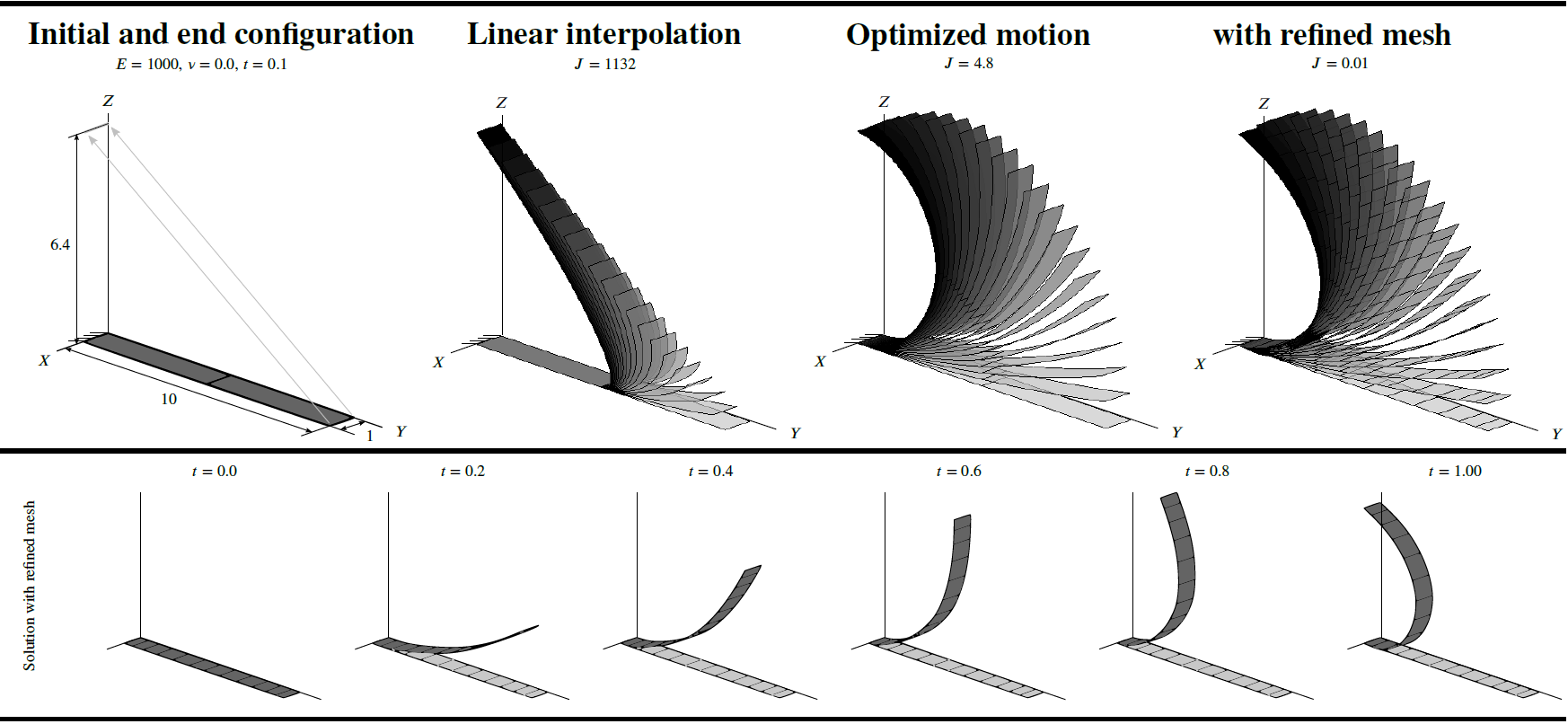}
\caption{Motion design of a cantilever with shell elements and corresponding inextensible deformations}
\label{fig:bsp_shell_cantilever}
\end{figure}

One important special case are inextensional deformations of developable structures with Gaussian curvature equal to zero, e.g. bending of a cylinder to a flat plane. The deformation of a cantilever beam illustrated in Figure~\ref{fig:bsp_shell_cantilever} represents the same phenomenon in a simple two-dimensional configuration. The left side is clamped and for the target configuration, the final location of the tip is prescribed. It is defined in a way that allows the final configuration to be a perfect half-circle.

Initially, the beam is discretized with only two quadratic isogeometric elements to improve convergence due to the low number of degrees of freedom. The path is discretized by two quadratic elements with B-spline shape functions. By motion design, an inextensional deformation is found, where the straight cantilever is bent to a half-circle while preserving its length. 

However, despite the good geometry approximation by using NURBS as shape functions, this mesh is too coarse to provide reasonable results in terms of stress and strain (and therefore the internal energy). Therefore, in the following an improved approximation of the geometry is realized by using 12 quadratic elements and the motion obtained with the coarse mesh is used as a predictor. The resulting motion is shown in Figure~\ref{fig:bsp_shell_cantilever} and it closely resembles the one obtained with the coarse mesh, but does not represent the perfect half circle. The value of the functional, however, significantly decreases from $J=4.8$ to $J=0.01$.

It has to be noted that in this numerical experiment the solution is not unique. Any deformed geometry for which the cantilever has the same length as the original flat configuration can be reached by an inextensional deformation. Accordingly, the problem is ill-posed. Nevertheless, one valid solution is found with apparently no numerical problems. In order to understand this surprising phenomenon one has to first understand that the finite elements used are based on a displacement based standard Galerkin formulation with no measures to avoid locking. For the problem at hand, \emph{membrane locking} is crucial. For the given discretization with 12 quadratic elements the effect is not very strong. However, the corresponding parasitic non-zero membrane strains are large enough to have a regularizing effect on the process of motion design.

\subsubsection{Transformation of a helicoid to a catenoid}

\begin{figure}[h]
\centering\small
\includegraphics[width=1.0\textwidth]{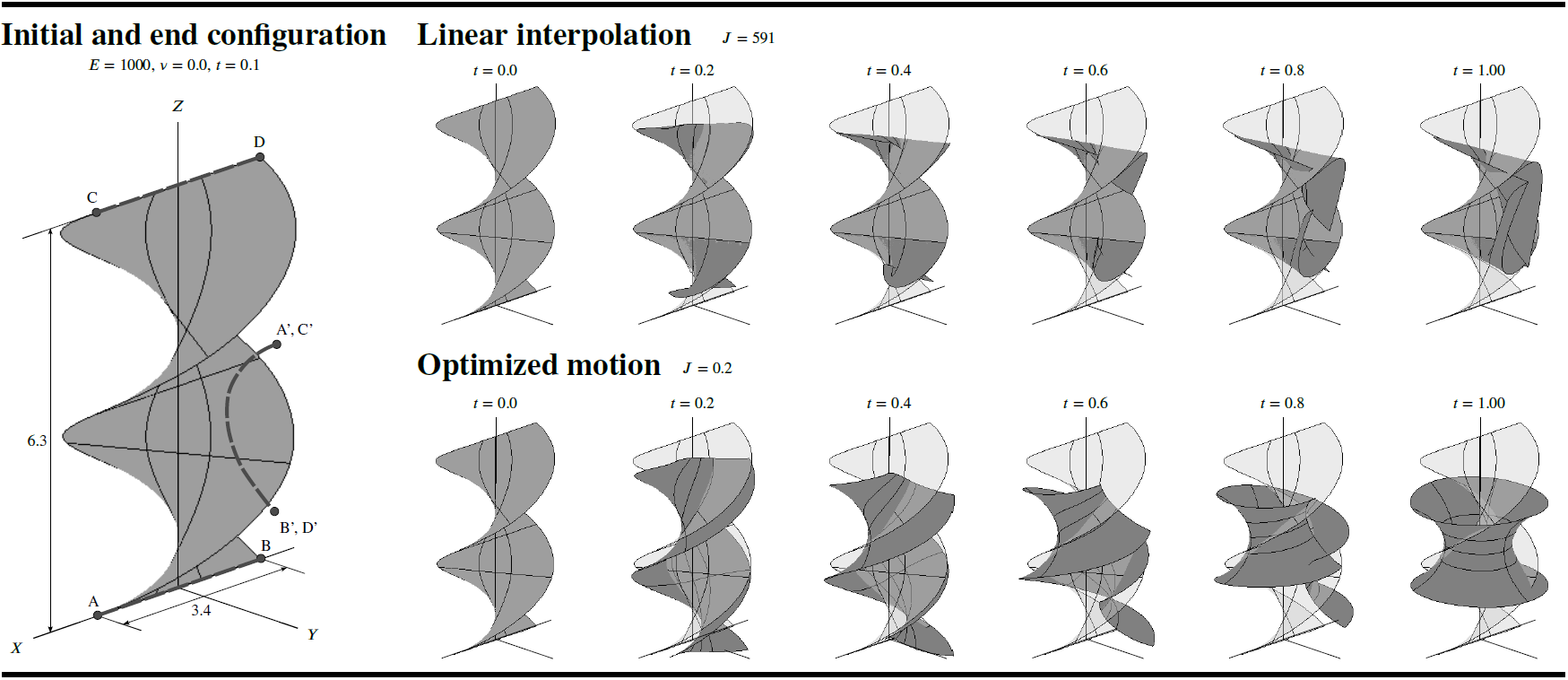}
\caption{Transformation from a helicoid to a catenoid with a motion design analysis}
\label{fig:bsp_shell_hel_cat}
\end{figure}

A classical example for an inextensional deformation is the transformation of a helicoid to a catenoid, shown in Figure~\ref{fig:bsp_shell_hel_cat}. It is a rare example from the field of analytical differential geometry for which an analytical solution for large inextensional deformations exists in the case of Gaussian curvature being non-zero.

The helicoid is discretized with $4 \times 4$ cubic elements with B-spline shape functions. For the target geometry, only the final position of the upper and lower Edge ($A-B$,$C-D$) are prescribed. The path is also coarsely discretized with two quadratic elements with B-spline shape functions. For motion design, the vertical displacement of a point at the upper edge is controlled.

Since the final geometry is only defined at the edges, the predictor, again obtained from a linear interpolation, shows a relatively bad first guess. The solution of the motion design problem not only determines the correct inextensible deformation, but also the correct final geometry, the catenoid. The value of the functional is $J=0.2$, i.e. again close to zero.

%%%%%%%%%%%%%%%%%%%%%%%%%%%%%%%%%%%%%%%%%%%%%%%%%%%%%%%%%%%%%

\section{Conclusions}
\label{sec:conclusions}

In this paper, a variational method for the design of motions of continuously deformable structures between two geometrical configurations, fulfilling certain desired properties and considering large displacements, has been presented. As a proof of concept, a functional defining the integrated internal energy along the motion path was defined. A combination of spatial discretization and path discretization is used for the numerical solution of the underlying problem.
The convergence behavior of the motion design problem is enhanced by various methods. In problems, where continuous deformation paths are expected, B-spline shape functions can be used to reduce the number of degrees of freedom compared to standard Lagrange discretization. An advantageous side effect of fewer degrees of freedom is a further improvement of convergence behavior in the iterative process.

The applied solution procedure can be interpreted as a second order optimization algorithm. For the problems studied herein, the derivatives (sensitivities) are calculated analytically. In the given framework this can be accomplished for any kind of finite element for spatial discretization.

Implementation of the motion design method was successfully verified by benchmark problems for which the exact solution is known. Additionally, the effect of instability and snap-through phenomena for motions with the prescribed functional was investigated in corresponding examples. The feasibility of the method for the design of inextensible deformations of shells was also demonstrated.

The successful application of motion design to detect or develop kinematic mechanisms reveals a genuine potential for application to adaptive and deployable structures.
The evolution of the required actuation forces during the deformation is recovered after the motion is found.
The restriction to those cases results from the fact, that the realization of the designed motion potentially requires forces at every degree of freedom, whereas usually, prescribed load cases or actuators are at hand. The future and subsequent steps in the development of the motion design method are therefore investigations on how to incorporate discrete actuator elements and that a resulting motion can be realized only with a restricted and specified amount of possible load cases.

%%%%%%%%%%%%%%%%%%%%%%%%%%%%%%%%%%%%%%%%%%%%%%%%%%%%%%%%%%%%%

\section*{Acknowledgements}
This work has been funded by the German Research Foundation (DFG) as part of the Transregional Collaborative Research Centre (SFB/Transregio) 141 “Biological Design and Integrative Structures”/project A04 under grant number INST 41/910-1 and the collaborative project ‘Bio-inspirierte Materialsysteme and Verbundkomponenten für nachhaltiges Bauen im 21ten Jahrhundert’ (BioElast), which is part of the ‘Zukunftsoffensive IV Innovation und Exzellenz – Aufbau und Stärkung der Forschungsinfrastruktur im Bereich der Mikro- und Nanotechnologie sowie der neuen Materialien’, funded by the State Ministry of Baden-Wuerttemberg for Sciences, Research and Arts.

%%%%%%%%%%%%%%%%%%%%%%%%%%%%%%%%%%%%%%%%%%%%%%%%%%%%%%%%%%%%%

\bibliographystyle{plain}
\bibliography{literature}

%%%%%%%%%%%%%%%%%%%%%%%%%%%%%%%%%%%%%%%%%%%%%%%%%%%%%%%%%%%%%

\appendix

\section{Exact solution of the Brachistochrone problem}
\label{app:brachistochrone_exact}

The starting point is eq.~(\ref{eq:brach_func}) in the main text, where the simplified Euler-Lagrange equation is applied for the functional of the Brachistochrone problem
\begin{align}
& \sqrt{\frac{1+y'^2}{2g(y_\RA - y)}} - y'\frac{y'}{\sqrt{2g(y_\RA-y)(1+y'^2)}}
= \frac{1}{\sqrt{2g(y_\RA-y)(1+y'^2)}} = C_1 .
\end{align}
Solving it for $y'$ yields
\begin{align}
y' = \sqrt{\frac{1}{2gC_1^2(y_\RA-y)}-1} =
\sqrt{\frac{1-2gC_1^2(y_\RA-y)}{2gC_1^2(y_\RA-y)}} .
\label{eq:brach_ex_ys}
\end{align}
A substitution is necessary for a solution and a clever choice for this problem is the parametric representation of trigonometrical functions
\begin{align}
2gC_1^2(y_\RA-y) = \sin^2\left(\frac{\bar{t}}{2}\right) = \frac 12 \big(1-\cos(\bar{t})\big)
\end{align}
and the resulting equation for $y$
\begin{align}
y = y_\RA - \frac{1}{2gC_1^2} \sin^2\left(\frac{\bar{t}}{2}\right) = y_\RA -
\frac{1}{4gC_1^2}\big(1-\cos(\bar{t})\big)
\end{align}
is now the ansatz. Inserting it into eq.~(\ref{eq:brach_ex_ys})
\begin{align}
y' = \sqrt{\frac{1-\sin^2\left(\frac{\bar{t}}{2}\right)}{\sin^2\left(\frac{\bar{t}}{2}\right)}} =
\frac{\cos\left(\frac{\bar{t}}{2}\right)}{\sin\left(\frac{\bar{t}}{2}\right)} = \fracdif{y}{x} 
\end{align}
enables the substitution
\begin{align}
\dx = \frac{\sin\left(\frac{\bar{t}}{2}\right)}{\cos\left(\frac{\bar{t}}{2}\right)} \dy .
\label{eq:dx}
\end{align}
The derivative of $y$ with respect to $\bar{t}$ is still necessary
\begin{align}
\fracdif{y}{\bar{t}} = \frac{1}{2gC_1^2} \sin \left(\frac{\bar{t}}{2}\right) \cos \left(\frac{\bar{t}}{2}\right) \qquad \rightarrow \qquad \dy = \frac{1}{2gC_1^2} \sin \left(\frac{\bar{t}}{2}\right) \cos \left(\frac{\bar{t}}{2}\right) \Rd \bar{t}
\end{align}
and can now be inserted into eq.~(\ref{eq:dx})
\begin{align}
\dx & = \frac{\sin\left(\frac{\bar{t}}{2}\right)}{\cos\left(\frac{\bar{t}}{2}\right)} \dy = \frac{\sin\left(\frac{\bar{t}}{2}\right)}{\cos\left(\frac{\bar{t}}{2}\right)}
\frac{1}{2gC_1^2} \sin \left(\frac{\bar{t}}{2}\right) \cos \left(\frac{\bar{t}}{2}\right) \Rd t = \frac{1}{2gC_1^2} \sin^2 \left(\frac{\bar{t}}{2}\right) \Rd t = \frac{1}{4gC_1^2} \big(1-\cos(\bar{t})\big)  \Rd \bar{t} .
\end{align}
By integration
\begin{align}
x & = \int \dx + C_2 = \int \frac{1}{4gC_1^2} \big(1-\cos(\bar{t})\big) \Rd \bar{t} + C_2 = \frac{1}{4gC_1^2} \left( \bar{t}- \sin (\bar{t}) \right) + C_2
\end{align}
the equations for $x$ and $y$ can be obtained
\begin{align}
y & = y_\RA - \frac{1}{2gC_1^2} \sin^2\left(\frac{\bar{t}}{2}\right) = y_\RA -
\frac{1}{4gC_1^2}\big(1-\cos(\bar{t})\big) \\
x & = \frac{1}{4gC_1^2} \left( \bar{t}- \sin (\bar{t}) \right) + C_2 .
\end{align}
As already explained in the main text, the constants $C_1$, $C_2$ as well as the parameter value $\bar{t}_\RE$ at point B can be derived by the boundary conditions of the starting point A and the end point B: $x(t=0) = x_\RA$, $x(\bar{t}=\bar{t}_\RE) = x_\RB$ and $y(\bar{t}=\bar{t}_\RE) = y_\RB$, which represent themselves non-linear functions that need to be solved iteratively. The condition $y(\bar{t}=0) = y_\RA$ is fulfilled by definition of the problem. Exemplary values are given for the starting point $x_\RA = 1.0$, $y_\RA = 5.0$ and the end point $x_\RB = 10.0$, $y_\RB = 2.0$ (calculated with the rounded gravitation constant $g = 10$)
\begin{align}
C_1 & = 0.116 &
C_2 & = 1.0 &
\bar{t}_\RE & = 4.05
\end{align}

%%%%%%%%%%%%%%%%%%%%%%%%%%%%%%%%%%%%%%%%%%%%%%%%%%%%%%%%%%%%%

\end{document}